\documentclass[iop]{emulateapj}

\usepackage{graphicx,amsmath}
\usepackage{color}
\usepackage[plainpages=false, colorlinks=true, anchorcolor=blue,
linkcolor=blue, citecolor=blue, bookmarks=false]{hyperref}

\newcommand{\rhomax}{\rho_{c}}

\newcommand{\Rmax}{R_{\rm A}}
\newcommand{\Rmin}{R_{\rm B}}

\newcommand{\rhosr}{\rho_{\rm sr}}
\newcommand{\hateta}{\widehat \eta}
\newcommand{\hatrho}{\widehat \rho}
\newcommand{\hatC}{\widehat {C}}
\newcommand{\hatPhi}{{\widehat \Phi}_s}
\newcommand{\hatomg}{\widehat \omega}
\newcommand{\hatomgmax}{{\widehat \omega}_{\rm max}}
\newcommand{\hatOmega}{{\widehat \Omega}_0}
\newcommand{\hatOmegas}{{\widehat \Omega}_s}
\newcommand{\hatOmegae}{{\widehat \Omega}_e}
\newcommand{\hatOmegac}{{\widehat \Omega}_{0,\rm crit}}
\newcommand{\hatr}{\widehat r}
\newcommand{\hatR}{\widehat R}
\newcommand{\hatz}{\widehat z}
\newcommand{\hatrB}{\widehat R_{\rm B}}
\newcommand{\hatrBmax}{\widehat R_{\rm B,1}}
\newcommand{\hatrBmin}{\widehat R_{\rm B,2}}

\newcommand{\hatM}{\widehat M}
\newcommand{\hatJ}{\widehat J}
\newcommand{\hatT}{\widehat T}
\newcommand{\hatW}{\widehat W}
\newcommand{\hatRmin}{\hatR_{\rm B}}
\newcommand{\tomega}{\omega_D}
\newcommand{\hatavgrho}{{\langle\widehat\rho\rangle}}
\newcommand{\avgrho}{{\langle\rho\rangle}}

\newcommand{\Msun}{{\rm\,M_\odot}}
\newcommand\kms{{\rm\, km\, s^{-1}}}
\newcommand\Mrate{{\Msun\rm\,yr^{-1}}}
\newcommand\pc{{\rm\, pc}}
\newcommand\kpc{{\rm\, kpc}}

\newcommand\cs{c_s}
\newcommand\etal{et al.~}

\newcommand\be{\begin{equation}}
\newcommand\ee{\end{equation}}
\newcommand\bea{\begin{eqnarray}}
\newcommand\eea{\end{eqnarray}}

\slugcomment{Accepted for publication in the ApJ}

\shorttitle{Gravitational Instability of Rings} %
\shortauthors{Kim \& Moon}

\begin{document}

\title{Equilibrium Sequences and Gravitational Instability of Rotating Isothermal Rings}

\author{Woong-Tae Kim$^{1,2}$ and Sanghyuk Moon$^1$}

\affil{$^1$Center for the Exploration of the Origin of the Universe
(CEOU), Department of Physics \& Astronomy, Seoul
National University, Seoul 151-742, Republic of Korea \\
$^2$Center for Theoretical Physics (CTP), Seoul National University, Seoul 151-742, Republic of Korea}

\begin{abstract}
Nuclear rings at centers of barred galaxies exhibit strong star
formation activities. They are thought to undergo gravitational
instability when sufficiently massive. We approximate them as
rigidly-rotating isothermal objects and investigate their gravitational instability. Using a self-consistent field method, we first construct their equilibrium sequences specified by two parameters: $\alpha$ corresponding to the thermal energy relative to gravitational potential energy, and $\hatrB$ measuring the ellipticity or ring thickness. Unlike in the incompressible case, not all values of $\hatrB$ yield an isothermal equilibrium, and the range of $\hatrB$ for  such equilibria shrinks with decreasing $\alpha$. The density distributions in the meridional plane are steeper for smaller $\alpha$, and well approximated by those of infinite cylinders for slender rings. We also calculate the dispersion relations of nonaxisymmetric modes in rigidly-rotating slender rings with angular frequency $\Omega_0$ and central density $\rhomax$. Rings with smaller $\alpha$ are found more unstable with a larger unstable range of the azimuthal mode number. The instability is completely suppressed by rotation when $\Omega_0$
exceeds the critical value. The critical angular frequency is found to be almost constant at $\sim 0.7 (G\rho_c)^{1/2}$ for $\alpha \gtrsim 0.01$ and increases rapidly for smaller $\alpha$. We apply our results to a sample of observed star-forming rings and confirm that rings without a noticeable azimuthal age gradient of young star clusters are indeed gravitationally unstable.
\end{abstract}


\section{Introduction}\label{s:intro}

Nuclear rings in barred-spiral galaxies often exhibit strong activities of star formation (e.g,
\citealt{but96,ken97,kna06,maz08,san10,maz11,hsi11,van11, oni15}). They are mostly circular, with ellipticity of $e\sim0-0.4$. They are thought to form due to nonlinear interactions of gas with an underlying non-axisymmetric stellar bar potential (e.g.,
\citealt{com85,but86,shl90,kna95,com01,com10}). Recent hydrodynamic
simulations show that the inflowing gas driven inward by the bar torque tends to gather at the location of centrifugal barrier, well inside the inner Lindblad resonance, where the centrifugal force on the gas balances the external gravity \citep{kim12b,kim12c,ks12,li15}. This predicts that nuclear rings are smaller in size in galaxies with stronger bars and/or lower pattern speeds, overall consistent with observational results of \citet{com10}.

One of important issues regarding nuclear rings is what determines the star formation rate (SFR) in them.  Observations indicate that the ring SFRs vary widely in the range of $\sim0.1$--$10\Mrate$ from galaxy to galaxy, with a smaller value corresponding to a more strongly barred galaxy \citep{maz08,com10}, although the total gas mass in each ring is almost constant at $\sim(1$--$6) \times 10^8\Msun$ (e.g., \citealt{but00,ben02,she05,sch06}).  By analyzing photometric H$\alpha$ data of 22 nuclear rings, \citet{maz08} found that about a half of their sample possesses an azimuthal age gradient of young star clusters in such a way that older ones are located systematically farther away from the contact points between a ring and dust lanes, while other rings do not show a noticeable age gradient (see also, e.g., \citealt{bok08,ryd10,bra12}).

To explain the spatial distributions of ages of young star clusters,
\citet{bok08} proposed two scenarios of star formation: the ``popcorn'' model in which star formation takes place in dense clumps randomly distributed along a nuclear ring, and the ``pearls on a string'' model where star formation occurs preferentially near the contact points. Since star clusters age as they orbit about the galaxy center, the pearls-on-a-string model naturally explains the presence of an azimuthal age gradient, while clusters with different ages are well mixed in the popcorn model (see also, e.g.,
\citealt{ryd01,ryd10,all06,san10,van13}). The most important factor
that determines the dominating type of star formation appears to be the mass inflow rate $\dot{M}$ to the ring along the dust lanes
\citep{seo13,seo14}. When $\dot{M}$ is less than a critical value
$\dot{M}_{c}$, all inflowing gas can be consumed at the contact points, and star formation occurs in the pearls-on-a-string fashion. When $\dot{M}> \dot{M}_{c}$, on the other hand, the inflowing gas overflows the contact points and is transferred into other parts of the ring, resulting in popcorn-style star formation when it becomes
gravitationally unstable. \citet{seo13} found numerically
$\dot{M}_c\sim 1\Mrate$ for typical nuclear rings, although it depends rather sensitively on the gas sound speed as well as the ring size.

The above consideration implicitly assumes that nuclear rings
undergoing star formation in the pearls-on-a-string manner are
gravitationally stable, while those with popcorn-type star formation
are globally unstable. However, this has yet to be tested
theoretically. Although several authors studied gravitational
instability of ring-like systems (e.g., \citealt{goo88,elm94,chr97,had11}), it is difficult to apply their
results directly to nuclear rings because of the approximations made in these studies. For example, \citet{goo88} analyzed a linear stability of shearing accretion rings (or tori) to gravitational perturbations, but their models were limited to incompressible gas without any motion along the vertical direction parallel to the rotation axis (see also \citealt{luy90,and97}). For magnetized compressible rings, \citet{elm94} showed that a ring with density larger than $0.6\kappa^2/G$ is gravitationally unstable, with $\kappa$ and $G$ referring to the epicycle frequency and the gravitational constant, respectively. However, this result was based on the local approximation that treated the ring as a thin uniform cylinder without considering its internal structure.

On the other hand, \citet{had11} and \citet{had14} analyzed stability
of polytropic rings with index $n=1.5$ by solving the linearized
equations as an initial value rather than eigenvalue problem, and found several unstable modes with the azimuthal mode number $m\leq 4$. \citet{chr97} instead ran two-dimensional nonlinear simulations of galaxy rings using the equations integrated along the vertical
direction. Using an adiabatic equation of state, they found that
massive slender rings are highly unstable to gravitating modes with $m$ as large as 18. However, these linear or nonlinear initial-value
approaches did not search all unstable modes systematically as functions of $m$ and rotation frequency $\Omega_0$.

Is a ring with given physical quantities (such as mass, size, sound
speed, rotation speed) gravitationally stable or not?  What is the most dominant mode if it is unstable? How fast does it grow? To address these questions, we in this paper perform a linear stability analysis of nuclear rings, assuming that they are slender and isothermal. We will find full dispersion relations of gravitationally unstable modes as well as the critical angular frequencies for stability. We will then apply the results to observed nuclear rings to check the presence or absence of an azimuthal age gradient of young star clusters is really consistent with stability properties of the rings. We will also run three-dimensional numerical simulations and compare the results with those of our linear stability analysis.

Stability analysis of any system requires to set up its initial
equilibrium a priori. Due to their complicated geometry, finding
equilibrium configurations of isothermal rings is a non-trivial task.
In a pioneering work, \citet{ost64b} treated the effects of rotation
and the curvature as perturbing forces to otherwise non-rotating
infinite cylinders, and obtained approximate expressions for density
distributions of polytropic or isothermal rings in axisymmetric
equilibrium.  To determine the equilibrium structure of a
slowly-rotating, spheroid-like body, \citet{ost68} developed a
self-consistent field (SCF) method that solves the Poisson equation as well as the equation for steady equilibrium, alternatively and
iteratively.  \citet{eri81} used a similar iteration method to find a
ring-like equilibrium sequence of incompressible bodies as a function
of $\Omega_0$. \citet{hac86a,hac86b} extended the original SCF method
of \citet{ost68} to make it work even for rapidly-rotating, ring-like
polytropes in two or three dimensions. In this paper, we shall modify
the SCF technique of \citet{hac86a} to find equilibrium sequences of
rigidly-rotating isothermal bodies. This will allow us to explore the
effects of compressibility on the internal structures of rings.

The remainder of this paper is organized as follows. In Section
\ref{s:eql}, we describe our SCF method used to construct isothermal
bodies in steady equilibrium. In Section \ref{s:seq}, we present the
equilibrium sequences of rigidly-rotating isothermal objects, together with test results for incompressible bodies and Bonner-Ebert spheres. We will also show that the density profiles of slender rings can well be approximated by those of infinite isothermal cylinders. In Section \ref{s:GI}, we perform a linear stability analysis of slender isothermal rings to obtain the dispersion relations as well as the critical angular frequencies, and present the results of numerical simulations. In Section \ref{s:sum}, we summarize and
conclude this work with applications to observed nuclear rings.

\section{SCF Method}\label{s:eql}

\subsection{Equilibrium Equations}

\begin{figure*}
\hspace{0.5cm}\includegraphics[angle=0, width=17cm]{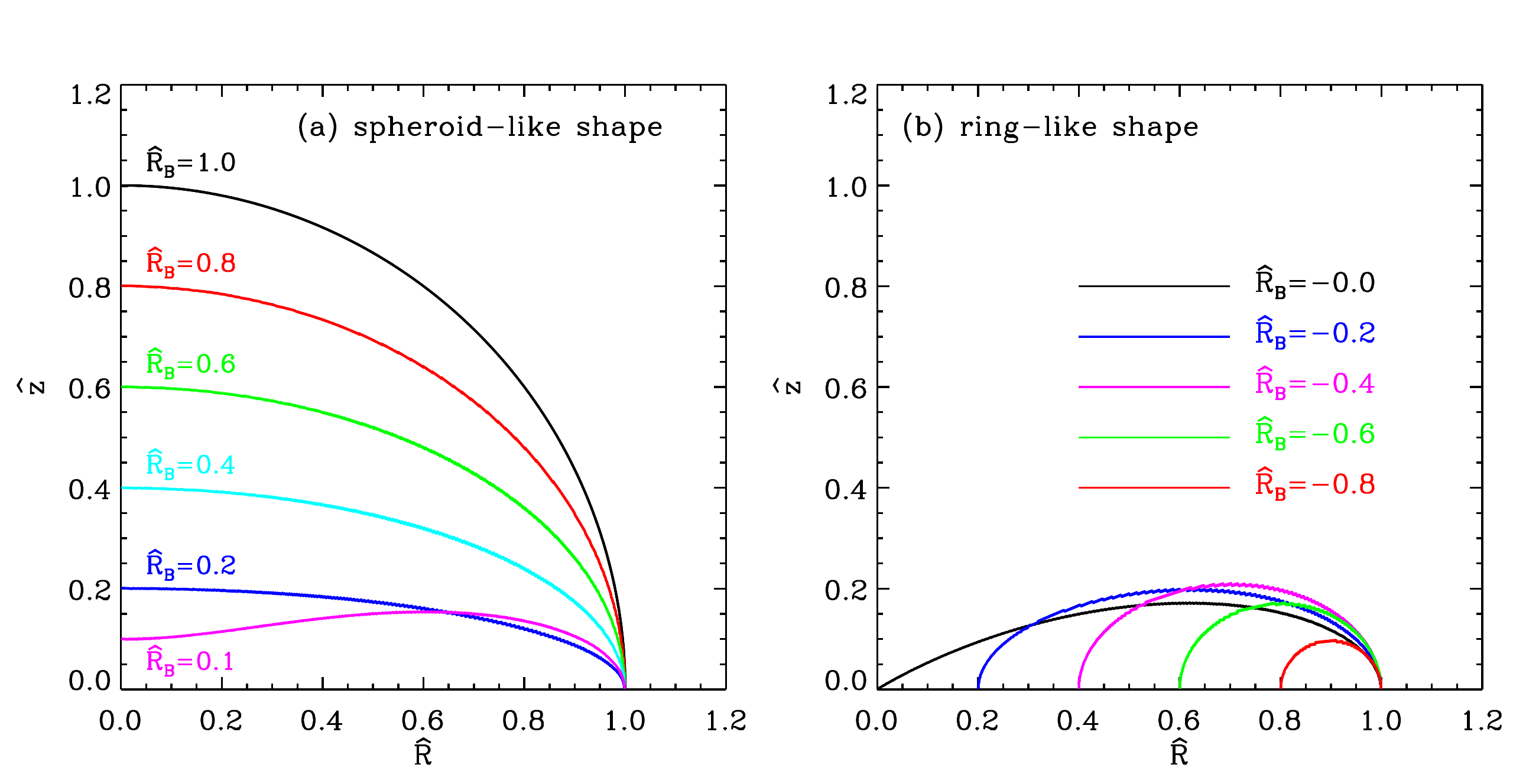}
 \caption{Shapes of the meridional cross section of axially-symmetric incompressible bodies in
  (a) spheroid-like configurations and (b) ring-like configurations.
 \label{f:incomp_shape}}
\end{figure*}

In this section, we explore equilibrium sequences of rotating,
isothermal bodies in the presence of both self-gravity and external
gravity. These bodies can take a spheroid-like or ring-like
configuration when the total angular momentum is small or large. We
assume that equilibrium bodies are rotating rigidly at angular
frequency $\Omega_0$ about its symmetry axis that is aligned in the
$\hat{z}$-direction. The equation of steady equilibrium then reads
 \be\label{e:HSE0}
   \cs^2 \nabla \ln \rho + \nabla \Phi_{\rm eff} = 0,
 \ee
where $\rho$ is the density, $\cs$ is the isothermal speed of sound,
and $\Phi_{\rm eff}$ is the  effective
potential defined by
 \be\label{e:effP}
  \Phi_{\rm eff} = \Phi_e + \Phi_s  - \frac{1}{2}\Omega_0^2 R^2,
 \ee
with $R$ being the cylindrical radial distance from the rotation axis. In Equation \eqref{e:effP}, $\Phi_e$ represents the external
gravitational potential, while $\Phi_s$ is the self-gravitational
potential satisfying the Poisson equation
  \be\label{e:pos0}
  \nabla^2\Phi_s = 4\pi G\rho.
 \ee
For nuclear rings in barred galaxies, $\Phi_e$ is provided mainly by a dark halo as well as a stellar disk and a bulge.  The last term in
Equation \eqref{e:effP} is the centrifugal potential.

We assume $\nabla \Phi_e = \Omega_e^2 \mathbf{R}$, so that the external gravity alone can make a body rotate at constant angular frequency $\Omega_e$. For nuclear rings, this approximation is valid if rings are geometrically thin. Equation \eqref{e:HSE0} is then integrated to yield
 \be\label{e:HSE1}
   \cs^2 \ln \rho +  \Phi_s  - \frac{1}{2}
   \Omega_s^2 R^2  = C,
 \ee
where $C$ is constant, and
  \be
    \Omega_s^2 \equiv \Omega_0^2 - \Omega_e^2.
  \ee
Note that $\Omega_s$ corresponds to the equilibrium angular frequency
of an \emph{isolated}, self-gravitating ring in the absence of the
external gravity.  Our aim is to obtain $\rho$ satisfying Equations
\eqref{e:pos0} and \eqref{e:HSE1} simultaneously.

To obtain equilibrium structure of slowly-rotating stars, \citet{ost68}
introduced an efficient SCF method that solves Equations \eqref{e:pos0}
and \eqref{e:HSE1} alternatively and iteratively. In the SCF method,
one first takes a trial distribution for $\rho$ and solves the Poisson equation to find $\Phi_s$, which in turn yields new $\rho$ from Equation \eqref{e:HSE1}. Calculations are repeated until the trial and new density distributions agree within a tolerance. \citet{hac86a} extended the original SCF method to make it suitable for rapidly rotating polytropes. Here, we closely follow Hachisu's method to determine isothermal equilibria.

Following \citet{hac86a}, we let $\rhomax$ and $\Rmax$ denote the
maximum density and the maximum radial extent in the equatorial plane
of an equilibrium object, respectively. We introduce the following
dimensionless variables:
 $ \hatrho \equiv {\rho}/{\rhomax},$
 $ \hatR \equiv {R}/{\Rmax},$
 $  \hatPhi \equiv {\Phi_s}/{(G\Rmax^2\rhomax)},$ and
 $  \hatOmegas \equiv {\Omega_s}/{(G\rhomax)^{1/2}}.$
Then, Equation \eqref{e:HSE1} reduces to
 \be\label{e:HSE2}
   \alpha \ln \hatrho = \hatC -  \hatPhi  + \frac{1}{2}
   \hatOmegas^2 \hatR^2,
 \ee
where $\hatC$ is a dimensionless constant and
 \be\label{e:alp}
    \alpha \equiv {\cs^2}/({G\Rmax^2\rhomax})
  \ee
measures the relative importance of the thermal to gravitational
potential energies.

In the SCF method, it is crucial to solve the Poisson equation accurately and efficiently. For spheroid-like configurations, it is customary to employ a multipole expansion technique on spherical polar coordinates. On the other hand, it is more efficient to utilize toroidal coordinates for ring-like configurations, especially for slender rings. In Appendix \ref{a:pairs}, we describe the methods to find $\Phi$ for given $\rho$ both in spherical and toroidal coordinates.

\subsection{Boundary Conditions}\label{s:bd}

In the case of a polytropic equation of state, an object in equilibrium achieves vanishing density at a finite radius, and is thus self-truncated. On the other hand, an isothermal object in steady equilibrium would extend to infinite distance without an external medium.  In reality, gas clouds or rings are usually in pressure equilibrium with their surrounding hot rarefied medium that provides a confining external pressure $P_{\rm ext}$. Fixing $P_{\rm ext}$ is equivalent to choosing $\hatC$ in Equation \eqref{e:HSE2}, or to placing the boundaries where $\rho=P_{\rm ext}/\cs^2$.

Following \citet{hac86a}, we let (positive) $\Rmin$ denote the radial
distance of the boundary along the $z$-axis for spheroid-like
configurations. For ring-like configurations, $\Rmin$ takes the
negative of the radial distance to the inner boundary in the equatorial plane. Then, Equation \eqref{e:HSE2} requires
 \be\label{e:Omg}
  \hatOmegas^2= \left\{\begin{array}{l}
    2[\hatPhi (1, \pi/2) - \hatPhi (\hatrB, 0)], \\
       \;\;\;\;\;\;\;\;\;\;\;\;\;\;\;\;\;\;\;\;\;\;\;\;\;\;\; \textrm{for spheroid-like bodies}, \\
    2[\hatPhi (1, \pi/2) - \hatPhi (-\hatrB, {\pi}/{2})]
  / {(1- \hatrB ^2)},
   \\ \;\;\;\;\;\;\;\;\;\;\;\;\;\;\;\;\;\;\;\;\;\;\;\;\;\;\;\textrm{for ring-like bodies},
  \end{array}\right.
 \ee
where
  \be\label{e:rB}
  \hatRmin \equiv \Rmin /\Rmax.
  \ee
with $0<\hatRmin\leq 1$ for spheroid-like bodies and $-1<\hatRmin<0$
for ring-like bodies. Since  $\hatOmegas$ depends on $\hatRmin$ through Equation \eqref{e:Omg}, an isothermal equilibrium can be completely specified by two parameters: $\alpha$ and $\hatRmin$.

Since $\Rmax$ is defined to be the maximum radial extent in the
meridional plane, the existence of an equilibrium demands
$\hatPhi-\hatOmegas^2\hatR^2/2$ in Equation \eqref{e:HSE2} to be an
increasing function of $\hatR$ near $\hatR=1$: the boundary should
otherwise retreat to a smaller radius where the thermal pressure is
equal to $P_{\rm ext}$. Since the potential minimum occurs inside
$\hatR=1$, this requires that the equilibrium should be sufficiently
self-gravitating and/or have small enough $\hatOmegas$. As will be
presented in Section \ref{s:iso},  an isothermal equilibrium turns out to be nonexistent for fairly small $|\Rmin|$ because
self-gravity is not strong enough or the angular frequency is too large to form gravitationally bound objects.

\subsection{Computation Method}

In Appendix \ref{a:comp}, we compare the results based on the potential expansions in the spherical and toroidal coordinates for ring-like equilibria, and show that the two methods agree with each other when $\hatrB\gtrsim -0.86$, while the multipole expansion in the spherical coordinates overestimates $\hatOmegas^2$ for smaller $\hatrB$. When we present the results in Section \ref{s:seq}, therefore, we employ the multipole expansion with $l_{\rm max}=10$ in the toroidal coordinates for flattened equilibria with $\hatrB\leq -0.8$, while adopting the spherical multipole expansion with $l_{\rm max}=50$ for any other equilibria.

As a domain of computation, we consider a meridional cross-section of
an equilibrium body, and divide it into $N_r\times N_a$ cells.
Here, $N_r$ and $N_a$ refer respectively to the mesh numbers over $0 \leq \hatr \leq 1.2$ in the $r$-direction and $0\leq \theta \leq \pi/2$ in the $\theta$-direction of the spherical coordinates, or
over $2.5\leq \sigma \leq 9.0$ in the $\sigma$-direction and over
$0\leq\tau\leq \pi$ in the $\tau$-direction of the toroidal coordinates.

Initially, we take $\hatrho= 1$ when $\hatr \leq 1$ and $\hatrho=0$ otherwise for spheroid-like configurations, and $\hatrho= 1$ when $\hatRmin \leq \hatr \sin\theta \leq 1$ and $\hatrho=0$ otherwise for ring-like configurations in spherical coordinates.  When we use the toroidal coordinates, we set the focal length equal to $\widehat{a}\equiv a/\Rmax = (1-\hatrB)/2$, and take $\hatrho=1$ in the regions with $(x-a)^2+z^2 \leq a^2$ and $\hatrho=0$ otherwise.
We then calculate $\rho_l$ from Equation
\eqref{e:exp1} or Equation \eqref{e:rho_to}, and $\Phi$ using Equation \eqref{e:pos1} or Equation \eqref{e:gpot_toroidal} based on
Gaussian and Newton-Cotes quadratures (e.g., \citealt{pre88}). Next, we calculate $\hatOmegas^2$ from Equation \eqref{e:Omg} and then update $\hatrho$ from Equation \eqref{e:HSE2}.  In each iteration on the toroidal mesh, $\widehat{a}$ is set to move to the location of the maximum density.  We repeat the calculations using the updated density until the relative difference in $\hatOmegas^2$ from two successive iterations is smaller than $10^{-6}$. We employ $N_r\times N_a$=$1024\times512$ cells, and it typically takes less than 20 iterations to obtain a converged solution.

Once we find an equilibrium configuration $\rho(r,\theta)$ in the spherical coordinates, it is straightforward to calculate its volume $V$, mass $M$, angular momentum $J$, kinetic energy $T$, and gravitational potential energy $W$ via
  \be
  M = 2\pi \int \rho r^2 \sin \theta dr d\theta,
  \ee
  \be
  J = 2\pi  \int \rho\Omega r^4 \sin^3 \theta dr d\theta,
  \ee
  \be
  T = \pi  \int \rho\Omega^2  r^4 \sin^3\theta dr d\theta,
  \ee
and
  \be
  W = \pi \int \rho  \Phi_s  r^2 \sin \theta dr d\theta.
  \ee
Note that $T=J\Omega/2$ for rigidly-rotating bodies. These quantities
will be evaluated and used to draw the equilibrium sequences of
isothermal rings in Section \ref{s:seq}.

\section{Equilibrium Configurations}\label{s:seq}

\subsection{Incompressible Bodies}\label{s:incomp}

\citet{eri81} found that an incompressible body in axial symmetry takes a form of a Maclaurin spheroid when rotating slowly, and bifurcates into a one-ring sequence when the total angular momentum exceeds a critical value (see also \citealt{cha67,bar71,hac86a}). As a test of our SCF method, we first apply it to construct equilibrium
configurations in the incompressible limit, which is attained by taking $\alpha\gg 1$. By comparing our results with \citet{eri81} for the Maclaurin spheroids and incompressible rings, we can check the accuracy of our SCF method.

\begin{figure}
\includegraphics[angle=0, width=8.5cm]{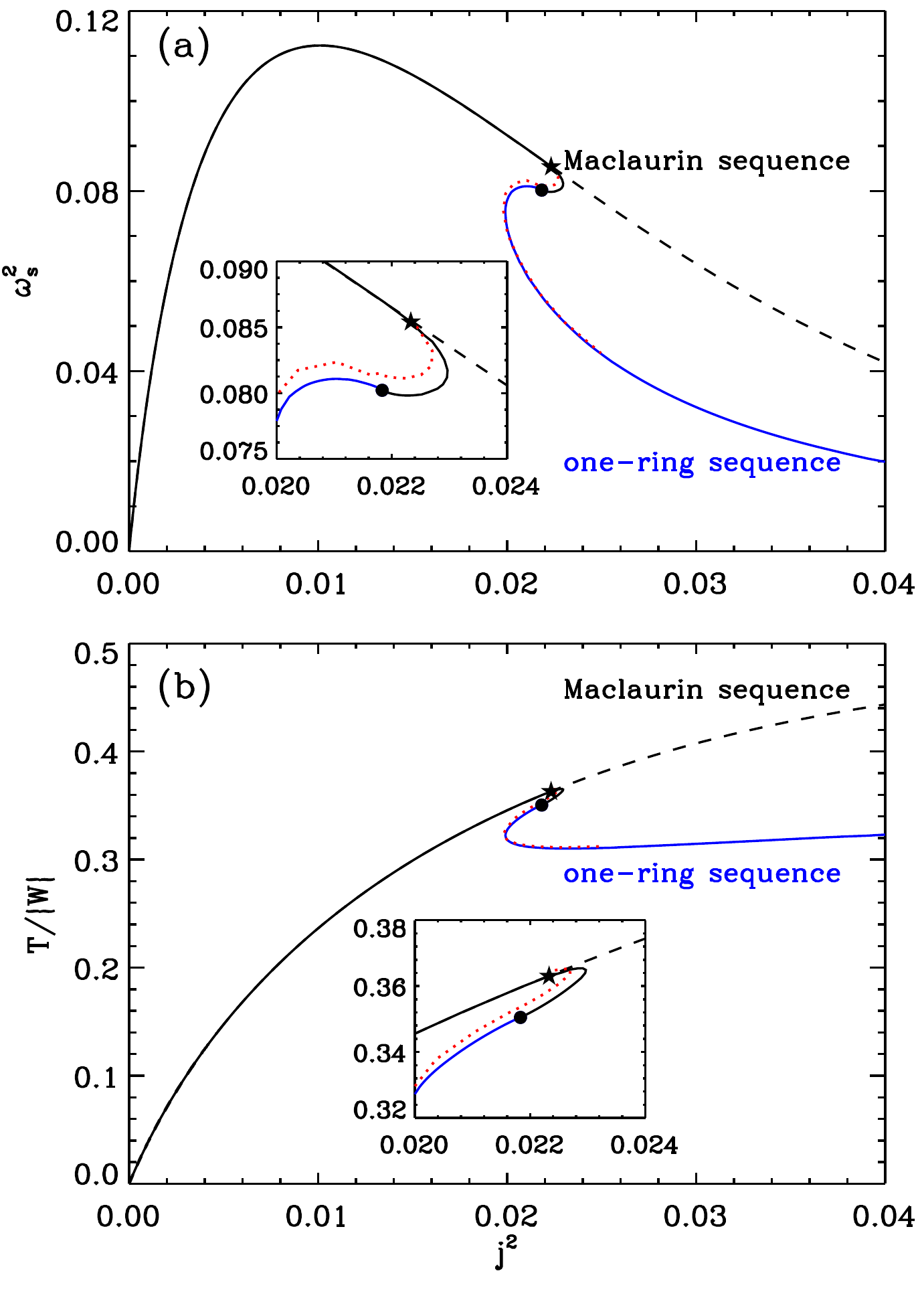}
 \caption{Dependence on the normalized angular momentum
 $j^2 = J^2/(4\pi G M^{10/3}\avgrho^{-1/3})$ of (a) the normalized
 angular velocity $\omega_s^2 = \Omega_s^2/(4\pi G \avgrho)$ and
 (b) the energy ratio $T/|W|$. The black and blue solid lines are our
 results with $\alpha=10^5$ for spheroid-like and ring-like
 equilibria, respectively. The black dashed lines correspond to
 the Maclaurin spheroid sequence, while
 the red dotted lines, adopted from Table 1 of \citet{eri81}, are
 for the hamburgers or the one-ring sequence.
 The filled stars at $j^2=2.233\times 10^{-2}$ indicate the bifurcation point from the Maclaurin sequence, while the open circles at
 $j^2=2.183\times 10^{-2}$ correspond to $\hatrB=0$.
 The insets zoom in the regions around the bifurcation point.
 \label{f:incomp_Omg}}
\end{figure}

\begin{deluxetable}{rlllll}
\tablecaption{Properties of Axially-symmetric Incompressible Bodies in Steady Equilibrium\label{t:incomp}}
\tablewidth{0pt} %
\tablehead{ \colhead{$\hatRmin$} & \colhead{$\hatOmegas^2$}
          & \colhead{$\hatM$}    & \colhead{$\hatJ$}
          & \colhead{$\hatT$}
          & \colhead{$-\hatW$} }
\startdata
   1.0 &    0.000E$+$0 &  4.189E$+$0 &   0.000E$+$0 &  0.000E$+$0 &   1.054E$+1$ \\
   0.9 &    3.263E$-$1 &  3.767E$+$0 &   8.599E$-$1 &  2.456E$-$1 &   8.811E$+0$ \\
   0.8 &    6.307E$-$1 &  3.349E$+$0 &   1.063E$+$0 &  4.221E$-$1 &   7.220E$+0$ \\
   0.7 &    9.082E$-$1 &  2.927E$+$0 &   1.115E$+$0 &  5.311E$-$1 &   5.729E$+0$ \\
   0.6 &    1.140E$+$0 &  2.510E$+$0 &   1.071E$+$0 &  5.718E$-$1 &   4.385E$+0$ \\
   0.5 &    1.314E$+$0 &  2.093E$+$0 &   9.581E$-$1 &  5.491E$-$1 &   3.179E$+0$ \\
   0.4 &    1.405E$+$0 &  1.671E$+$0 &   7.910E$-$1 &  4.688E$-$1 &   2.121E$+0$ \\
   0.3 &    1.380E$+$0 &  1.254E$+$0 &   5.880E$-$1 &  3.454E$-$1 &   1.253E$+0$ \\
   0.2 &    1.192E$+$0 &  8.389E$-$1 &   3.660E$-$1 &  1.998E$-$1 &   5.902E$-1$ \\
   0.1 &    1.014E$+$0 &  7.885E$-$1 &   3.609E$-$1 &  1.817E$-$1 &   5.012E$-1$ \\
   0.0 &    1.008E$+$0 &  8.504E$-$1 &   3.998E$-$1 &  2.007E$-$1 &   5.726E$-1$ \\
$-0.1$ &    1.019E$+$0 &  8.796E$-$1 &   4.162E$-$1 &  2.100E$-$1 &   6.103E$-1$ \\
$-0.2$ &    9.845E$-$1 &  9.403E$-$1 &   4.528E$-$1 &  2.246E$-$1 &   6.844E$-1$ \\
$-0.3$ &    8.743E$-$1 &  9.397E$-$1 &   4.537E$-$1 &  2.121E$-$1 &   6.703E$-1$ \\
$-0.4$ &    7.189E$-$1 &  8.615E$-$1 &   4.076E$-$1 &  1.728E$-$1 &   5.558E$-1$ \\
$-0.5$ &    5.459E$-$1 &  7.182E$-$1 &   3.235E$-$1 &  1.195E$-$1 &   3.845E$-1$ \\
$-0.6$ &    3.794E$-$1 &  5.366E$-$1 &   2.214E$-$1 &  6.818E$-$2 &   2.162E$-1$ \\
$-0.7$ &    2.321E$-$1 &  3.439E$-$1 &   1.224E$-$1 &  2.947E$-$2 &   9.076E$-2$ \\
$-0.8$ &    1.130E$-$1 &  1.696E$-$1 &   4.661E$-$2 &  7.835E$-$3 &   2.311E$-2$ \\
$-0.9$ &    3.204E$-$2 &  4.654E$-$2 &   7.537E$-$3 &  6.745E$-$4 &
 1.911E$-3$
\enddata
 \tablecomments{The number behind E indicates the exponent of the
  power of 10.}
\end{deluxetable}

Figure \ref{f:incomp_shape} plots the boundaries in the meridional
plane of axially-symmetric incompressible bodies in (a) spheroid-like
and (b) ring-like equilibrium for some selected values of $\hatRmin$.
For all cases, $\alpha=10^5$ is taken. For $0.158 \lesssim \hatRmin
\leq 1$, an equilibrium is exactly a Maclaurin spheroid with an
ellipticity $e = (1-\hatRmin^2)^{1/2}$. The cases with $0 < \hatRmin
\lesssim 0.158$ result in concave ``hamburgers'' that are somewhat more flared at intermediate $\hatR$ $(\sim 0.6-0.8)$ than at the symmetry axis (e.g., \citealt{eri81,hac86a}).  When $\hatRmin <0$, on the other hand, equilibrium bodies take a form of rotating rings (or tori), with a larger $|\hatRmin|$ corresponding to a more slender ring. Table \ref{t:incomp} lists the dimensionless quantities $\hatOmegas$, $\hatM = M/(\Rmax^3\rhomax)$, $\hatJ = J/(G^{1/2}\Rmax^5\rhomax^{3/2})$, $\hatT = T/(G\Rmax^5\rhomax^2)$, and $\hatW = W/(G\Rmax^5\rhomax^2)$ for incompressible bodies.

\begin{figure}
\includegraphics[angle=0, width=8.5cm]{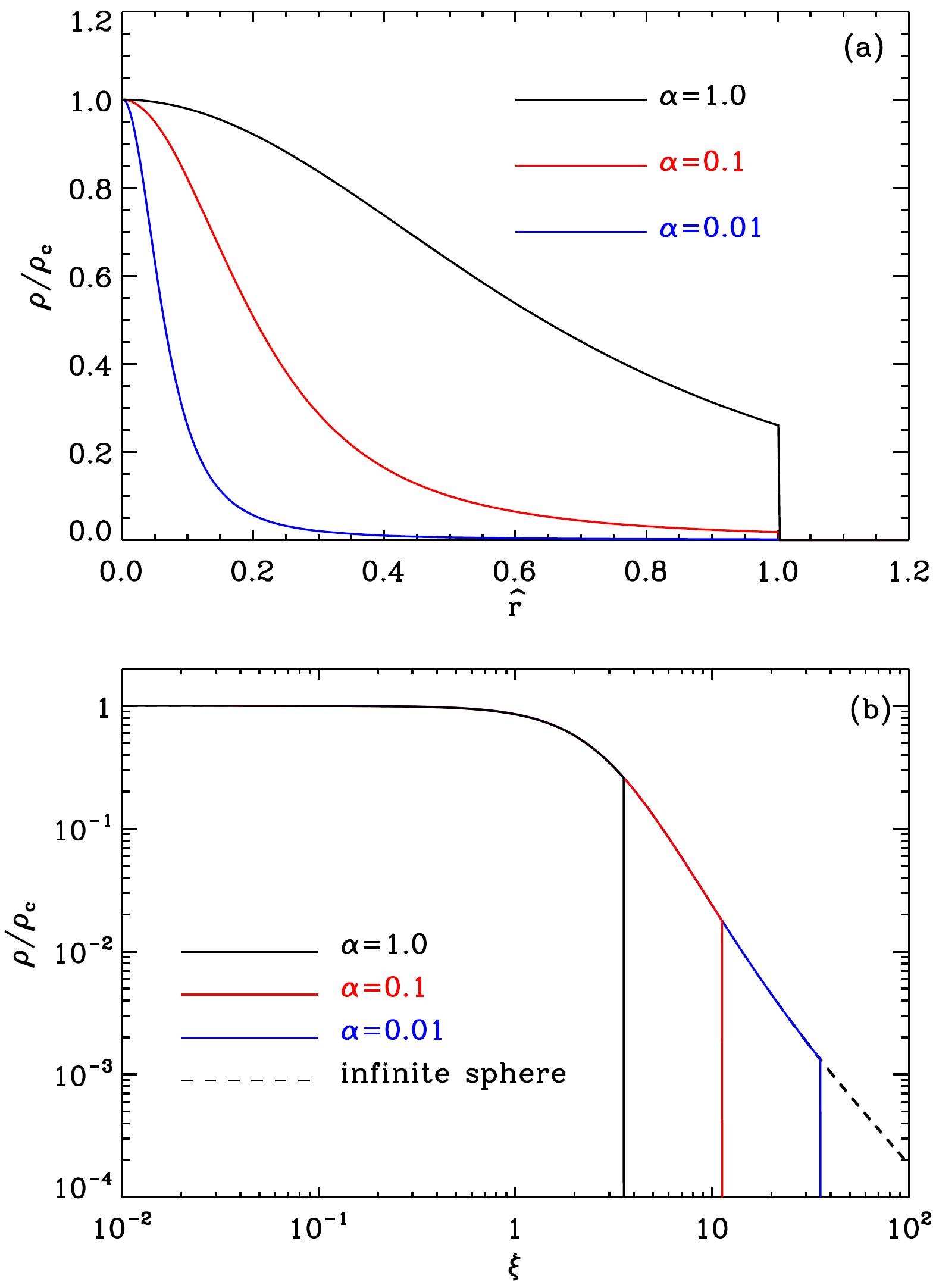}
 \caption{ (a) Density distributions of non-rotating isothermal spheres as functions of the dimensionless radius (a) $\hatr$ and (b)
 $\xi = (4\pi/\alpha)^{1/2} \hatr$ for
 $\alpha=1$, $0.1$, and $0.01$.
 All the cases are truncated at $\hatr=1$.
 The dashed line in (b) represents
 an infinite isothermal sphere
 (without pressure truncation).
  Note that $\rho$ is independent of $\alpha$ except for the truncation radius. Smaller $\alpha$ corresponds to truncation at
  smaller external pressure.
 \label{f:isosphere}}
 \vspace{0.2cm}
\end{figure}

\begin{figure*}
\hspace{0.5cm}\includegraphics[angle=0, width=17.cm]{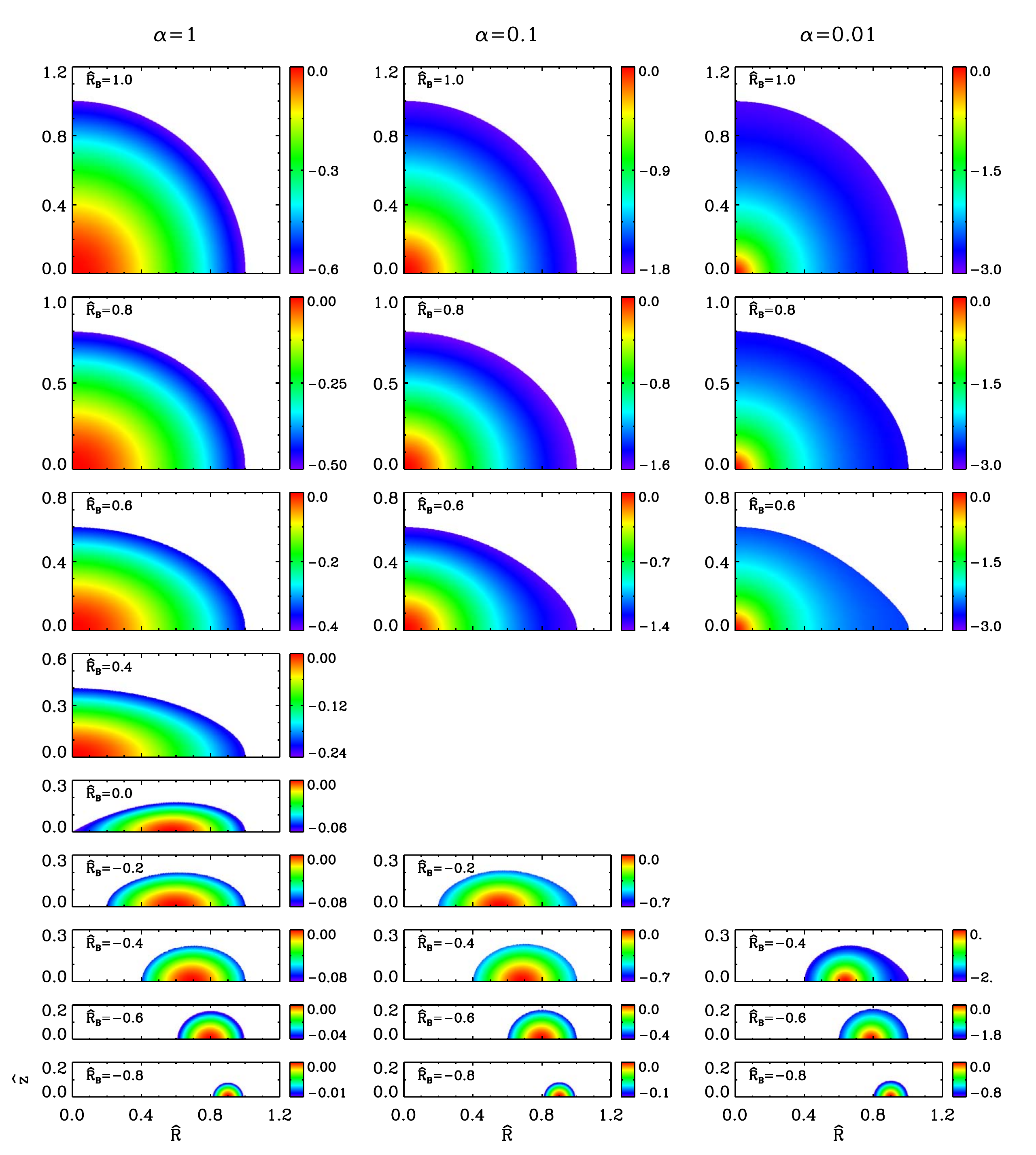}
 \caption{Equilibrium density distributions on the meridional plane of isothermal bodies with
 $\Rmin=1.0$, 0.8, 0.6, 0.4, 0.0, $-0.2$, $-0.4$, $-0.6$, and $-0.8$ from top to bottom.
 The left, middle, and right columns correspond to $\alpha=1$, 0.1, and 0.01,
respectively. Colorbars label $\log \rho/\rhomax$.
 \label{f:isocontour}}
\end{figure*}

Figure \ref{f:incomp_Omg} plots (a) the square of the normalized
angular velocity $\omega_s^2 \equiv \Omega_s^2/(4\pi G\avgrho)$ and (b) the ratio of the kinetic to gravitational potential energy $t\equiv T/|W|$ as functions of the square of the normalized angular momentum $j^2\equiv J^2/(4\pi G M^{10/3} \avgrho^{-1/3})$. Here, $\avgrho$ denotes the volume-averaged density. The black and blue solid lines are the spheroid-like and ring-like equilibria, respectively, that we obtain by taking $\alpha=10^5$. The black dashed lines plot the theoretical predictions
 \be
   \omega_s^2= \frac{(1-e^2)^{1/2}}{2e^2}
    \left[(3-2e^2)\frac{\sin^{-1}e}{e} - 3(1-e^2)^{1/2}\right],
 \ee
 \be
    t = \frac{3}{2e^2} - 1 - \frac{3(1-e^2)^{1/2}}{2e\sin^{-1}e},
 \ee
and
 \be
    j^2 = \frac{4 \omega_s^2}{25} \left(\frac{3}{4\pi}\right)^{4/3}
     (1-e^2)^{-2/3},
 \ee
of the Maclaurin spheroid sequence (e.g., Eqs.~(3.234), (3.236), and
(3.239) of \citealt{lan99}), which are in excellent agreement with our results at the small-$j$ part. The filled stars at
$j^2=2.233\times10^{-2}$ (occurring at $\hatRmin=0.158$) mark the
bifurcation point where the Maclaurin sequence branches out into the
concave hamburgers and then into the one-ring sequence after the filled circles at $j^2=2.183\times 10^{-2}$ (or $\hatRmin=0$).  The insets zoom in the regions with $0.020 \leq j^2 \leq 0.024$ around the bifurcation point for clarity. For comparison, we also plot the
$\omega_s^2$--$j^2$ and $T/|W|$--$j^2$ relationships adopted from Table 1 of \citet{eri81} as the red dotted lines, which are slightly
($\sim1$--$2$\%) different from our results (see also \citealt{hac86a}). These discrepancies are presumably due to the
insufficient resolution used by these authors in solving the Poisson
equation.\footnote{\citet{hac86a} employed $N_r\times N_a$=$257\times277$ cells and truncated the multipole expansion at $l_{\rm max}=16$ in his SCF calculations.}

\begin{figure*}[!t]
\hspace{0.5cm}\includegraphics[angle=0, width=17cm]{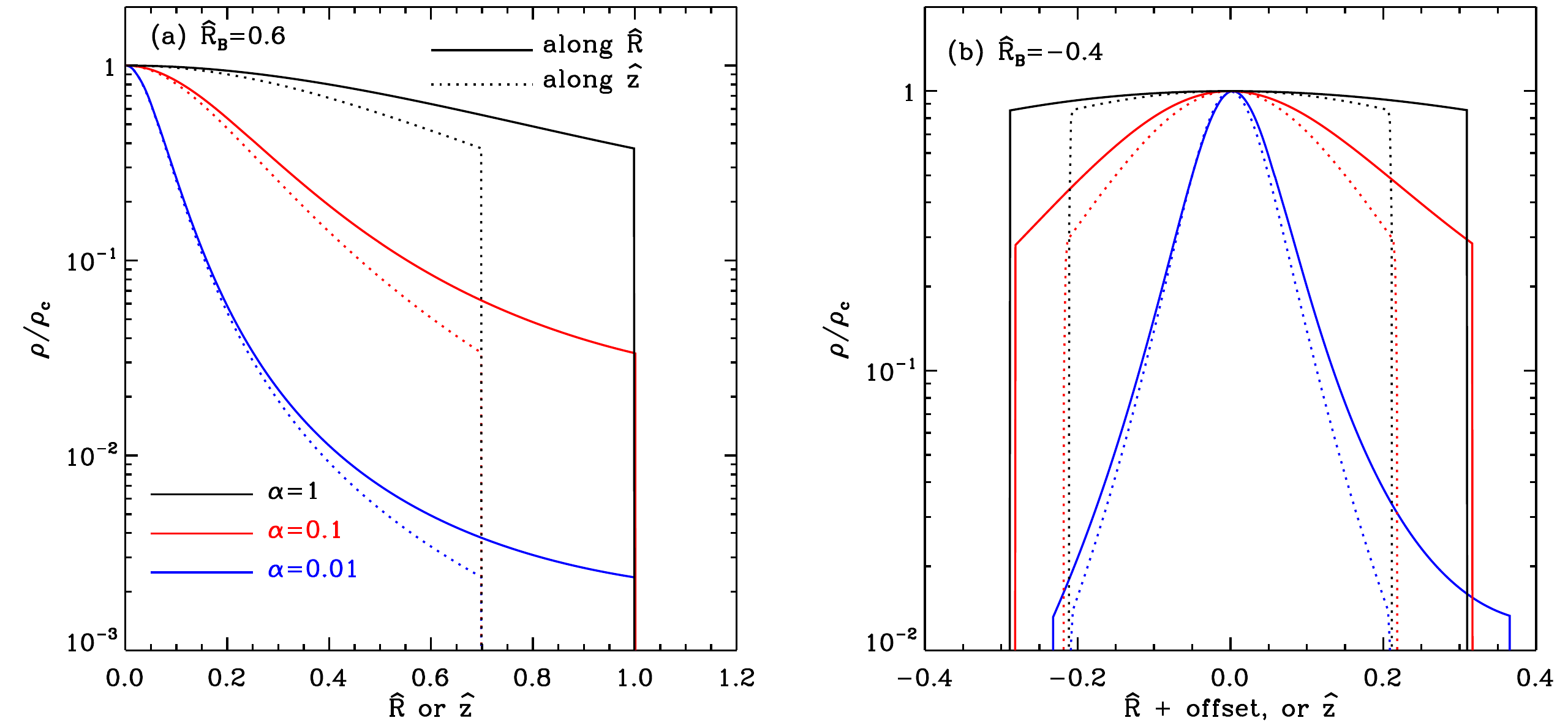}
 \caption{Equilibrium density profiles of (a) spheroid-like objects with $\hatRmin=0.6$ and (b) ring-like objects with $\hatRmin=-0.4$. The cases with $\alpha=1$, 0.1, and 0.01 are shown as black, red, and blue curves, respectively.  The solid and dotted lines are along the $\hatR$- and $\hatz$-axis, respectively. In (b), the solid curves are shifted horizontally to make the maximum density occur at zero in the abscissa.
 \label{f:isoprofile}}
\end{figure*}

\subsection{Isothermal Objects}\label{s:iso}

\begin{figure*}
\hspace{0.5cm}\includegraphics[angle=0, width=17cm]{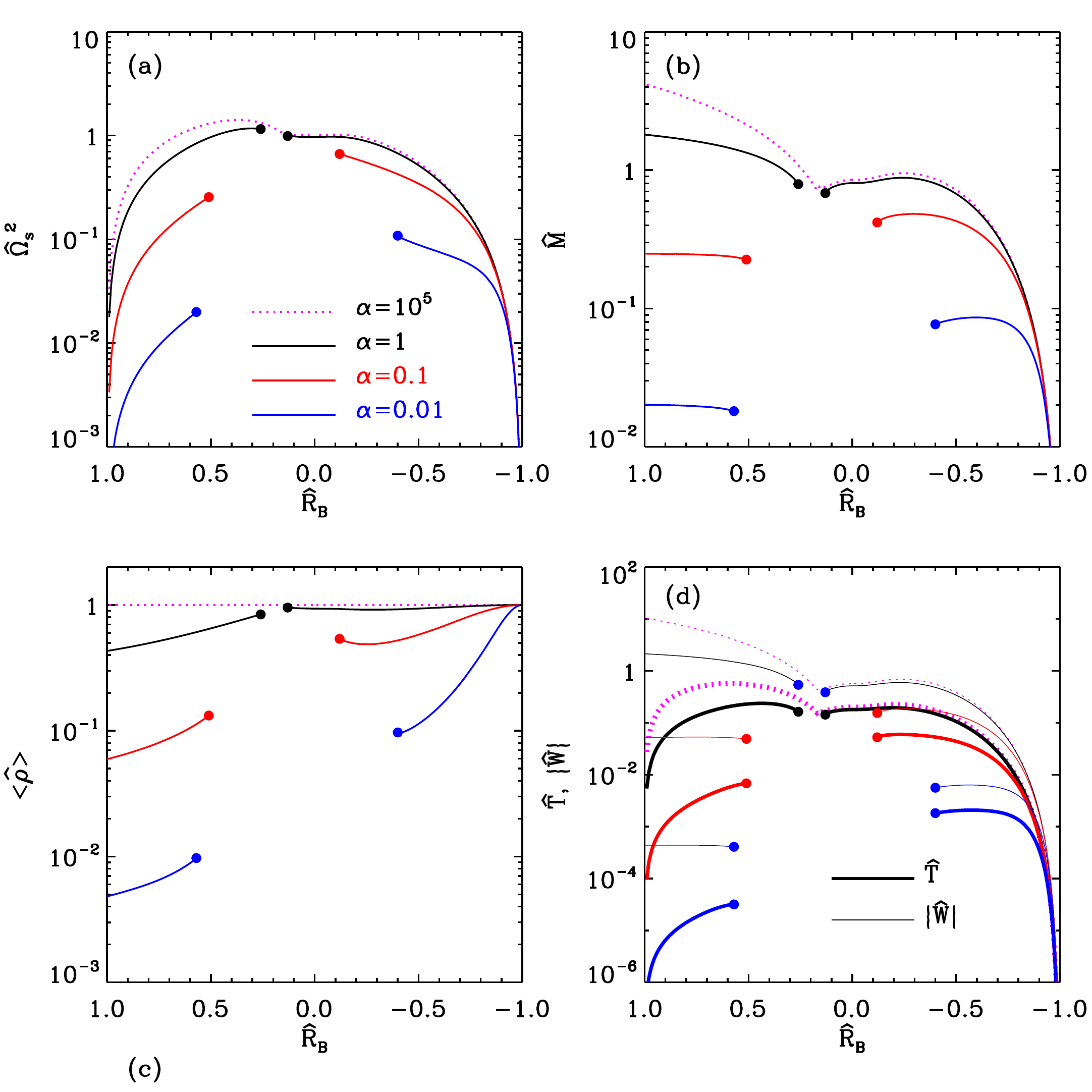}
 \caption{Dependence on $\hatRmin$ of (a) the angular frequency $\hatOmegas^2$, (b)
the total mass $\hatM$, (c) the average density $\hatavgrho$, and (d) the kinetic energy $\hatT$ (thick lines) and the gravitational potential energy $|\hatW|$ (thin lines) for isothermal equilibria with $\alpha=1$, 0.1, and 0.01. Filled circles mark the ranges of $\Rmin$ for the existence of isothermal equilibria. The incompressible cases with $\alpha=10^5$ are compared as dotted lines.
 \label{f:isoRB}}
\end{figure*}

We now present density distributions of isothermal objects in
axisymmetric equilibrium. We first visit non-rotating isothermal
spheres truncated by an external pressure. We then explore how rotation changes equilibrium structures.

\subsubsection{Bonnor-Ebert Spheres}

\begin{deluxetable*}{rllllll}
\tabletypesize{\small}
\tablecaption{Properties of Axially-symmetric Isothermal Bodies in Steady Equilibrium\label{t:iso}}
\tablewidth{0pt} %
\tablehead{ \colhead{$\hatRmin$} & \colhead{$\hatOmegas^2$}
          & \colhead{$\hatM$}    & \colhead{$\hatavgrho$}    & \colhead{$\hatJ$}
          & \colhead{$\hatT$}
          & \colhead{$-\hatW$} }
\startdata
          &               &                & $\alpha=1$   &               &               &      \\
\hline
     1.0  &   0.000E$+$0  &   1.806E$+$0  &   4.310E$-$1  &   0.000E$+$0  &   0.000E$+$0  &   2.140E$+$0   \\
     0.9  &   1.861E$-$1  &   1.735E$+$0  &   4.611E$-$1  &   2.566E$-$1  &   5.536E$-$2  &   2.028E$+$0   \\
     0.8  &   3.805E$-$1  &   1.657E$+$0  &   4.964E$-$1  &   3.551E$-$1  &   1.095E$-$1  &   1.900E$+$0   \\
     0.7  &   5.826E$-$1  &   1.564E$+$0  &   5.392E$-$1  &   4.209E$-$1  &   1.606E$-$1  &   1.746E$+$0   \\
     0.6  &   7.810E$-$1  &   1.455E$+$0  &   5.901E$-$1  &   4.599E$-$1  &   2.032E$-$1  &   1.561E$+$0   \\
     0.5  &   9.653E$-$1  &   1.322E$+$0  &   6.512E$-$1  &   4.709E$-$1  &   2.313E$-$1  &   1.335E$+$0   \\
     0.4  &   1.111E$+$0  &   1.152E$+$0  &   7.243E$-$1  &   4.453E$-$1  &   2.347E$-$1  &   1.054E$+$0   \\
     0.3  &   1.171E$+$0  &   9.247E$-$1  &   8.065E$-$1  &   3.672E$-$1  &   1.987E$-$1  &   7.133E$-$1   \\
     0.1  &   9.806E$-$1  &   7.472E$-$1  &   9.465E$-$1  &   3.316E$-$1  &   1.642E$-$1  &   4.531E$-$1   \\
     0.0  &   9.681E$-$1  &   8.052E$-$1  &   9.350E$-$1  &   3.662E$-$1  &   1.802E$-$1  &   5.162E$-$1   \\
  $-$0.1  &   9.728E$-$1  &   8.281E$-$1  &   9.305E$-$1  &   3.781E$-$1  &   1.865E$-$1  &   5.440E$-$1   \\
  $-$0.2  &   9.272E$-$1  &   8.744E$-$1  &   9.202E$-$1  &   4.044E$-$1  &   1.947E$-$1  &   5.950E$-$1   \\
  $-$0.3  &   8.176E$-$1  &   8.696E$-$1  &   9.173E$-$1  &   4.030E$-$1  &   1.822E$-$1  &   5.767E$-$1   \\
  $-$0.4  &   6.726E$-$1  &   7.999E$-$1  &   9.232E$-$1  &   3.644E$-$1  &   1.494E$-$1  &   4.811E$-$1   \\
  $-$0.5  &   5.156E$-$1  &   6.745E$-$1  &   9.360E$-$1  &   2.946E$-$1  &   1.058E$-$1  &   3.402E$-$1   \\
  $-$0.6  &   3.633E$-$1  &   5.119E$-$1  &   9.523E$-$1  &   2.064E$-$1  &   6.222E$-$2  &   1.971E$-$1   \\
  $-$0.7  &   2.257E$-$1  &   3.338E$-$1  &   9.700E$-$1  &   1.171E$-$1  &   2.782E$-$2  &   8.563E$-$2   \\
  $-$0.8  &   1.115E$-$1  &   1.671E$-$1  &   9.854E$-$1  &   4.562E$-$2  &   7.617E$-$3  &   2.246E$-$2   \\
  $-$0.9  &   3.193E$-$2  &   4.636E$-$2  &   9.961E$-$1  &   7.494E$-$3  &   6.695E$-$4  &   1.897E$-$3   \\

\hline
          &               &                & $\alpha=0.1$ &               &               &      \\
\hline
     1.0  &   0.000E$+$0  &   2.490E$-$1  &   5.941E$-$2  &   0.000E$+$0  &   0.000E$+$0  &   5.219E$-$2   \\
     0.9  &   3.749E$-$2  &   2.480E$-$1  &   6.607E$-$2  &   1.141E$-$2  &   1.104E$-$3  &   5.258E$-$2   \\
     0.8  &   8.080E$-$2  &   2.467E$-$1  &   7.483E$-$2  &   1.696E$-$2  &   2.410E$-$3  &   5.292E$-$2   \\
     0.7  &   1.319E$-$1  &   2.440E$-$1  &   8.718E$-$2  &   2.177E$-$2  &   3.954E$-$3  &   5.295E$-$2   \\
     0.6  &   1.912E$-$1  &   2.385E$-$1  &   1.053E$-$1  &   2.573E$-$2  &   5.625E$-$3  &   5.212E$-$2   \\
  $-$0.2  &   5.916E$-$1  &   4.694E$-$1  &   4.925E$-$1  &   1.546E$-$1  &   5.944E$-$2  &   1.831E$-$1   \\
  $-$0.3  &   5.060E$-$1  &   4.825E$-$1  &   4.910E$-$1  &   1.645E$-$1  &   5.852E$-$2  &   1.861E$-$1   \\
  $-$0.4  &   4.242E$-$1  &   4.684E$-$1  &   5.224E$-$1  &   1.632E$-$1  &   5.316E$-$2  &   1.709E$-$1   \\
  $-$0.5  &   3.439E$-$1  &   4.281E$-$1  &   5.802E$-$1  &   1.499E$-$1  &   4.394E$-$2  &   1.407E$-$1   \\
  $-$0.6  &   2.632E$-$1  &   3.596E$-$1  &   6.605E$-$1  &   1.224E$-$1  &   3.141E$-$2  &   9.898E$-$2   \\
  $-$0.7  &   1.809E$-$1  &   2.632E$-$1  &   7.617E$-$1  &   8.246E$-$2  &   1.754E$-$2  &   5.379E$-$2   \\
  $-$0.8  &   9.929E$-$2  &   1.478E$-$1  &   8.706E$-$1  &   3.804E$-$2  &   5.993E$-$3  &   1.764E$-$2   \\
  $-$0.9  &   3.093E$-$2  &   4.481E$-$2  &   9.627E$-$1  &   7.130E$-$3  &   6.270E$-$4  &   1.774E$-$3   \\

\hline
          &               &               &$\alpha=0.01$  &               &               &      \\
\hline
     1.0  &   0.000E$+$0  &   2.018E$-$2  &   4.831E$-$3  &   0.000E$+$0  &   0.000E$+$0  &   4.390E$-$4   \\
     0.9  &   3.318E$-$3  &   2.006E$-$2  &   5.349E$-$3  &   2.255E$-$4  &   6.494E$-$6  &   4.400E$-$4   \\
     0.8  &   7.235E$-$3  &   1.984E$-$2  &   6.058E$-$3  &   3.313E$-$4  &   1.409E$-$5  &   4.391E$-$4   \\
     0.7  &   1.199E$-$2  &   1.946E$-$2  &   7.109E$-$3  &   4.155E$-$4  &   2.275E$-$5  &   4.344E$-$4   \\
     0.6  &   1.779E$-$2  &   1.862E$-$2  &   8.857E$-$3  &   4.593E$-$4  &   3.064E$-$5  &   4.188E$-$4   \\
  $-$0.4  &   1.084E$-$1  &   7.682E$-$2  &   9.693E$-$2  &   1.105E$-$2  &   1.819E$-$3  &   5.609E$-$3   \\
  $-$0.5  &   8.838E$-$2  &   8.399E$-$2  &   1.134E$-$1  &   1.363E$-$2  &   2.027E$-$3  &   6.225E$-$3   \\
  $-$0.6  &   7.534E$-$2  &   8.618E$-$2  &   1.551E$-$1  &   1.509E$-$2  &   2.071E$-$3  &   6.293E$-$3   \\
  $-$0.7  &   6.356E$-$2  &   8.293E$-$2  &   2.362E$-$1  &   1.517E$-$2  &   1.913E$-$3  &   5.714E$-$3   \\
  $-$0.8  &   4.852E$-$2  &   6.856E$-$2  &   4.028E$-$1  &   1.229E$-$2  &   1.354E$-$3  &   3.937E$-$3   \\
  $-$0.9  &   2.391E$-$2  &   3.357E$-$2  &   7.206E$-$1  &   4.664E$-$3  &   3.583E$-$4  &   1.005E$-$3
\enddata
 \tablecomments{The number behind E indicates the exponent of the
  power of 10.}
\end{deluxetable*}

Consider non-rotating, self-gravitating isothermal spheres with
$\Omega_s=0$, namely Bonner-Ebert spheres, in hydrostatic equilibrium. Equations \eqref{e:pos0} and \eqref{e:HSE1} are then combined to yield
 \be\label{e:LE}
   \frac{1}{\xi^2} \frac{d}{d\xi}
   \left(\xi^2\frac{d\psi}{d\xi}\right) = \exp(-\psi),
 \ee
where $\psi=(\Phi_s-C)/\cs^2$ and $\xi= \left({4\pi
G\rhomax}/{\cs^2}\right)^{1/2} r = \left({4\pi}/{\alpha}\right)^{1/2}
\hatr$ are the dimensionless potential and radius, respectively.
Equation \eqref{e:LE} can be solved subject to the proper boundary
conditions, $\psi=d\psi/d\xi=0$ at $\xi=0$, to give the density
distribution $\rho=\rhomax \exp({-\psi})$, which shall be compared with the results of our SCF method.

As a second test, we apply our SCF method to obtain density
distributions of isothermal spheres by setting $\hatRmin=1$. Figure
\ref{f:isosphere}(a) plots the resulting density profiles for
$\alpha=1$, 0.1, and 0.01 as functions of $\hatr$: all the cases are
truncated at $\hatr=1$. Note that $\rho$ varies more steeply for
smaller $\alpha$ in order to compensate for a smaller sound speed in
balancing self-gravity. Note also that when drawn against $\xi$, as
shown in Figure \ref{f:isosphere}(b), all the curves lie well (within
$1\%$) on the inner parts of the solution of Equation \eqref{e:LE}
plotted as a dashed line, confirming the accuracy of our SCF method. A smaller $\alpha$ corresponds to a larger truncation radius in $\xi$.

\subsubsection{Rigidly-rotating Isothermal Equilibria}

\begin{figure}
\includegraphics[angle=0, width=8.5cm]{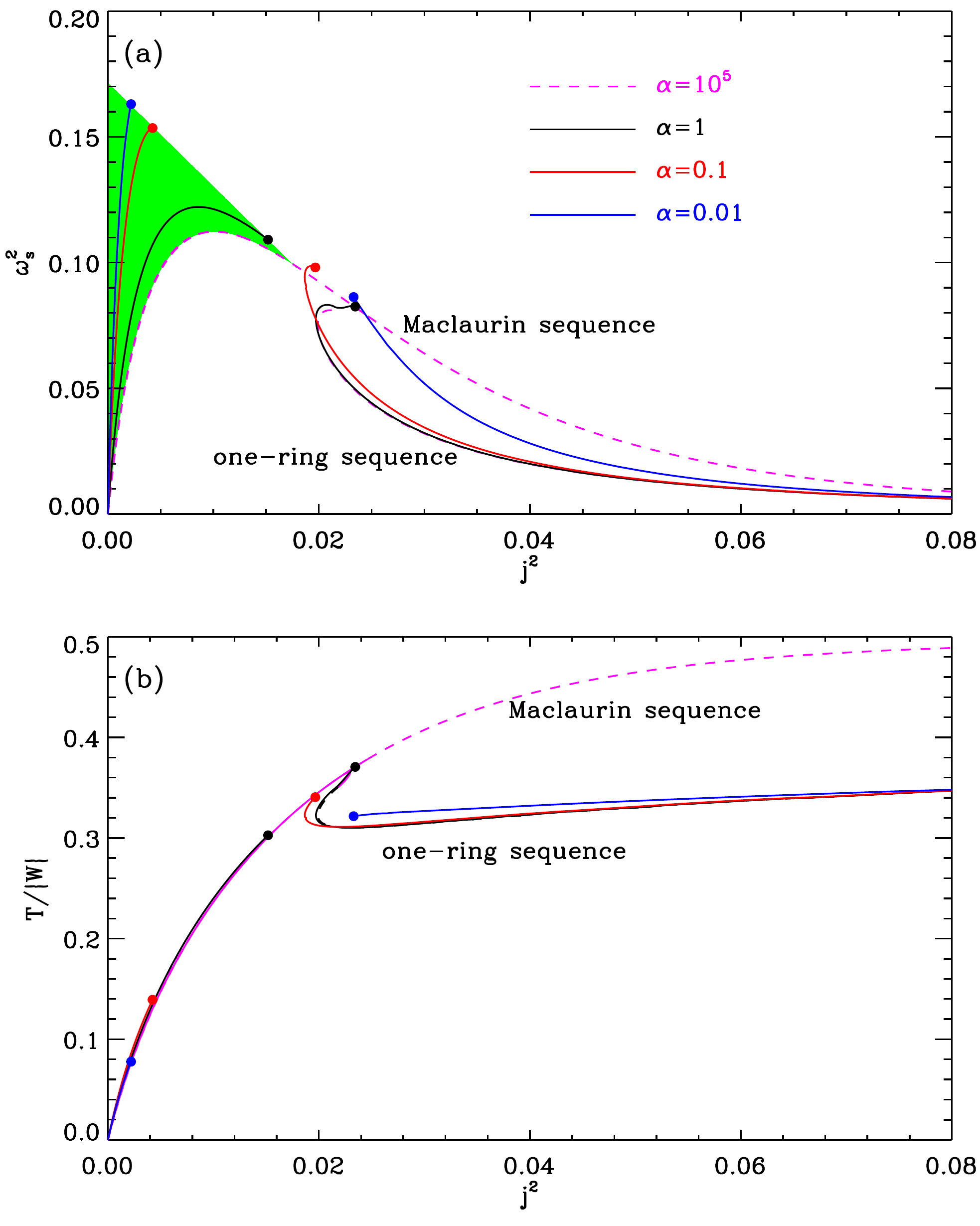}
 \caption{Dependence on the angular momentum $j^2$ of (a) the
 angular velocity $\omega_s^2$ and
 (b) the energy ratio $T/|W|$ for isothermal equilibria with $\alpha=1$, 0.1, and 0.01.
 The green shade in (a) represents the regions where
 spheroid-like isothermal equilibria can exist.
 \label{f:isoJ}}
\end{figure}

\begin{figure*}
\includegraphics[angle=0, width=18cm]{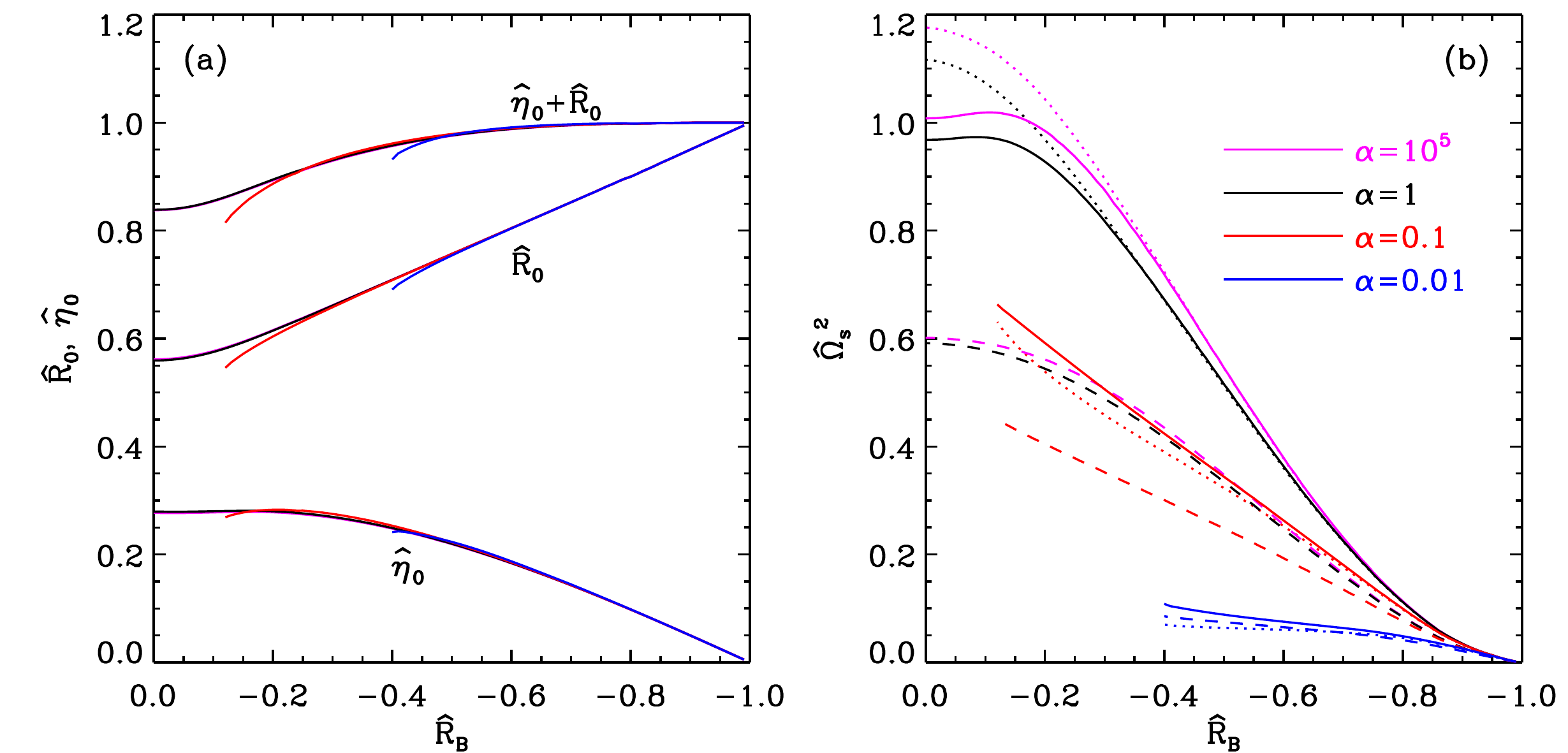}
\caption{Dependence on $\hatRmin$ of (a) the major axis $\hatR_0$ and
minor axis $\hateta_0$, and (b) the angular frequency $\hatOmegas^2$ of ring-like equilibria for $\alpha=10^5$ (purple), 1 (black), 0.1 (red), and 0.01 (blue).  Note that $\hatR_0$ and $\hateta_0$ for $\alpha=1$ are almost identical to those with $\alpha=10^5$. In (b), the solid lines give the results of our SCF method, while the dotted and dashed lines draw the analytic expressions of \citet{ost64b} for
incompressible and infinitely-extended isothermal rings, respectively.
 \label{f:sleOmg}}
\end{figure*}

By taking $\hatRmin$ less than unity, we obtain equilibrium density of isothermal objects in rigid rotation.  Figure \ref{f:isocontour}
presents the resulting density distributions on the meridional plane
for such equilibria with differing $\hatRmin$. The left, middle, and
right columns are for $\alpha=1$, 0.1, and 0.01, respectively. Figure
\ref{f:isoprofile} plots the exemplary density profiles along the
$\hatR$-axis (solid lines) and the $\hatz$-axis (dotted lines) for the spheroid-like configurations with $\hatRmin=0.6$ and the ring-like configurations with $\hatRmin=-0.4$. Clearly, an equilibrium is more centrally concentrated for smaller $\alpha$. The vertical extent of an equilibrium body is smaller than the horizontal extent for both spheroid-like and ring-like objects. Unlike Bonner-Ebert spheres whose density distributions are independent of $\alpha$ when expressed in terms of $\xi$, we find that the density profiles of rotating isothermal objects with different $\alpha$ along the $\hatz$- or $\hatR$-axis cannot be expressed by a single function of $\left({4\pi}/{\alpha}\right)^{1/2} \hatz$ or
$\left({4\pi}/{\alpha}\right)^{1/2} \hatR$.

Figure \ref{f:isoRB} plots the variations of the square of the angular velocity $\hatOmegas^2$, the total mass $\hatM$, the averaged density $\hatavgrho$, the kinetic energy $\hatT$, and the gravitational potential energy $\hatW$ as functions of $\hatRmin$ for isothermal equilibria with $\alpha=1$, 0.1, and 0.01, in comparison with the incompressible cases. Both $\hatOmegas$ and $\hatM$ increase as $|\hatRmin|$ decreases from unity.  For spheroid-like configurations, this is because an equilibrium body becomes more flattened and occupies a smaller volume as it rotates faster. On the other hand, ring-like configurations attain a larger volume and mass with decreasing $|\hatRmin|$, and thus requires larger $\hatOmegas$ to balance self-gravity.  Note, however, that $\hatM$ is not a monotonically decreasing function of $\hatRmin$ for ring-like configurations due to complicated dependence of their volume on $\hatRmin$ (e.g., Fig.~\ref{f:incomp_shape}b). For $\alpha=0.1$, for example, $\hatM$ increases as $\hatRmin$ moves away from $-1$, is maximized at $\hatRmin=-0.30$, and starts to decreases afterwards. Obviously, $\hatavgrho=1$ for incompressible configurations due to uniform density. Overall, $\hatavgrho$ increases with decreasing $\hatRmin$ and tends to unity as $\hatRmin$ approaches $-1$. The dependency of $\hatT$ and $\hatW$ on $\hatRmin$ is closely related to that of $\hatOmegas^2$ and $\hatM$, respectively.  For $\alpha \lesssim 0.1$, $\hatM$ and $\hatW$ are insensitive to $\hatRmin\gtrsim0.6$ since the density in the outer parts is very small. All the quantities are smaller with smaller $\alpha$ due to a stronger density concentration: these values are also listed in Table \ref{t:iso} for some selected values of $\hatRmin$.

The dependencies of $\hatOmegas$ and $\hatM$ on $\hatRmin$ make an
equilibrium sequence with fixed $\alpha\;(\leq 1)$ cease to exist for
an intermediate range of $\hatRmin$: $0.13 < \hatRmin < 0.27 $ for
$\alpha=1$, $-0.14 <\hatRmin < 0.51$ for $\alpha=0.1$, and $-0.40 <
\hatRmin < 0.59$ for $\alpha=0.01$, with the corresponding boundaries
marked by filled circles in Figure \ref{f:isoRB}. This is unlike the
incompressible bodies for which any value of $0\leq |\hatRmin| \leq 1$ readily yields an equilibrium.  As mentioned in Section \ref{s:bd}, the presence of a steady equilibrium requires large enough $|\hatPhi|$ and/or small enough $\hatOmegas$ in order for $\rho$ to be a decreasing function of $\hatR$ near $\hatR=1$ (see Eq.~\eqref{e:HSE2}). The absence of an isothermal equilibrium for intermediate $\hatRmin$ results from the fact that the self-gravitational potential is too weak to overcome the centrifugal potential in establishing gravitationally bound objects.

Figure \ref{f:isoJ} plots  the square of the normalized angular
velocity $\omega_s^2$ as well as  the energy ratio $t=T/|W|$ as
functions of the square of the normalized angular momentum $j^2$ for
isothermal equilibria with $\alpha=1$, 0.1, and 0.01. The
incompressible cases with $\alpha=10^5$ are compared as dashed lines.
Comparison of Figure \ref{f:isoJ} with Figures 10 and 11 of
\citet{hac86a} reveals that the $\omega_s^2$--$j^2$ and $t$--$j^2$
relationships of isothermal objects with $\alpha \sim 0.01$--1 are very close to those of polytropes with index $n\sim 0.1$--1.5.
\citet{hac86a} found that spheroid-like polytropic equilibria can be
possible only in the region $\omega_s^2+5j^2<0.185$ when $j^2<0.02$.
Our results suggest that spheroid-like isothermal equilibria can exist in the shaded region in Figure \ref{f:isoJ}(a), which is bounded by
 \be
  \omega_s^2+ 4.12 j^2 = 0.172,
 \ee
and the incompressible $\omega_s^2$--$j^2$ relation.

\subsection{Slender Rings}

Here we focus on the properties of slender isothermal rings whose minor axis is much shorter than the major axis. Density distributions of such rings can be obtained by taking $\hatRmin$ close to $-1$ in our SCF method. Using a perturbation analysis, \citet{ost64b} derived
approximate expressions for the density and angular frequency of both
polytropic and isothermal rings in steady equilibrium. Our objective in this subsection is to compare the results of our SCF method with
\citet{ost64b}.

For a ring-like configuration with $\hatRmin<0$, we define its major
axis $R_0$ and minor axis $\eta_0$ as
 \be\label{e:axis}
   R_0 = {\mathcal V}/(2\pi {\mathcal A}),\;\;\;\text{and}
   \;\;\;\eta_0=({\mathcal A}/\pi)^{1/2},
 \ee
where ${\mathcal V}$ and ${\mathcal A}$ refers to the total volume and the meridional cross section, respectively, occupied by the equilibrium body.  We plot in Figure \ref{f:sleOmg}(a) as solid lines
$\hatR_0=R_0/\Rmax$ and $\hateta_0= \eta_0/\Rmax$ together with
$\hatR_0+\hateta_0$ resulting from the SCF method as functions of
$\hatRmin$ for $\alpha=10^5$, 1, 0.1, and 0.01. Note that $\hateta_0
\simeq (1+\hatRmin)/2$ and $\hatR_0 \simeq (1-\hatRmin)/2$, resulting
in $\hatR_0 + \hateta_0 \simeq 1$, as expected, for $\hatRmin \lesssim -0.6$. This indicates that $R_0$ and $\eta_0$ defined in Equation \eqref{e:axis} describe the real major and minor axes of slender rings reasonably well.

Using a perturbation analysis, \citet{ost64b} showed that an
incompressible, slender ring in the absence of external gravity should obey
 \be\label{e:Ostinc}
   \hatOmegas^2= \frac{M}{2\pi R_0^3\rhomax} \left[
     \ln \left(\frac{8R_0}{\eta_0}\right) - \frac{5}{4} \right].
 \ee
For an isothermal ring of infinite extent (without pressure
truncation), he also derived
 \be\label{e:Ostiso}
   \hatOmegas^2= \frac{M_\infty}{2\pi R_0^3\rhomax} \left[
     \ln \left(\frac{8R_0}{\eta_{1/2}}\right) - 2 \right],
 \ee
which is valid when $\eta_{1/2}/R_0\ll1$. Here, $M_\infty$ denotes the total mass and $\eta_{1/2}=(M_\infty/2\pi^2R_0 \rhomax)^{1/2}$ is the half-mass radius. Figure \ref{f:sleOmg}(b) plots Equation
\eqref{e:Ostinc} as dotted lines, which are in good agreement with the results of our SCF method, shown as solid lines, for $\alpha \gtrsim 1$ at $\hatRmin \lesssim -0.4$ and even for $\alpha$ as small as 0.01 at $\hatRmin\lesssim -0.8$. Figure \ref{f:sleOmg}(b) also plots Equation \eqref{e:Ostiso} as dashed lines after taking $M=M_\infty$, which matches well our SCF results only for $\alpha=0.01$. The discrepancy between Equation \eqref{e:Ostiso} and our results is in fact expected since isothermal rings considered in the present paper are truncated by external pressure. Since rings with $\alpha=0.01$ are highly concentrated, however, Equation \eqref{e:Ostiso} can still be a good approximation for truncated slender rings.

\begin{figure}
\includegraphics[angle=0, width=8.5cm]{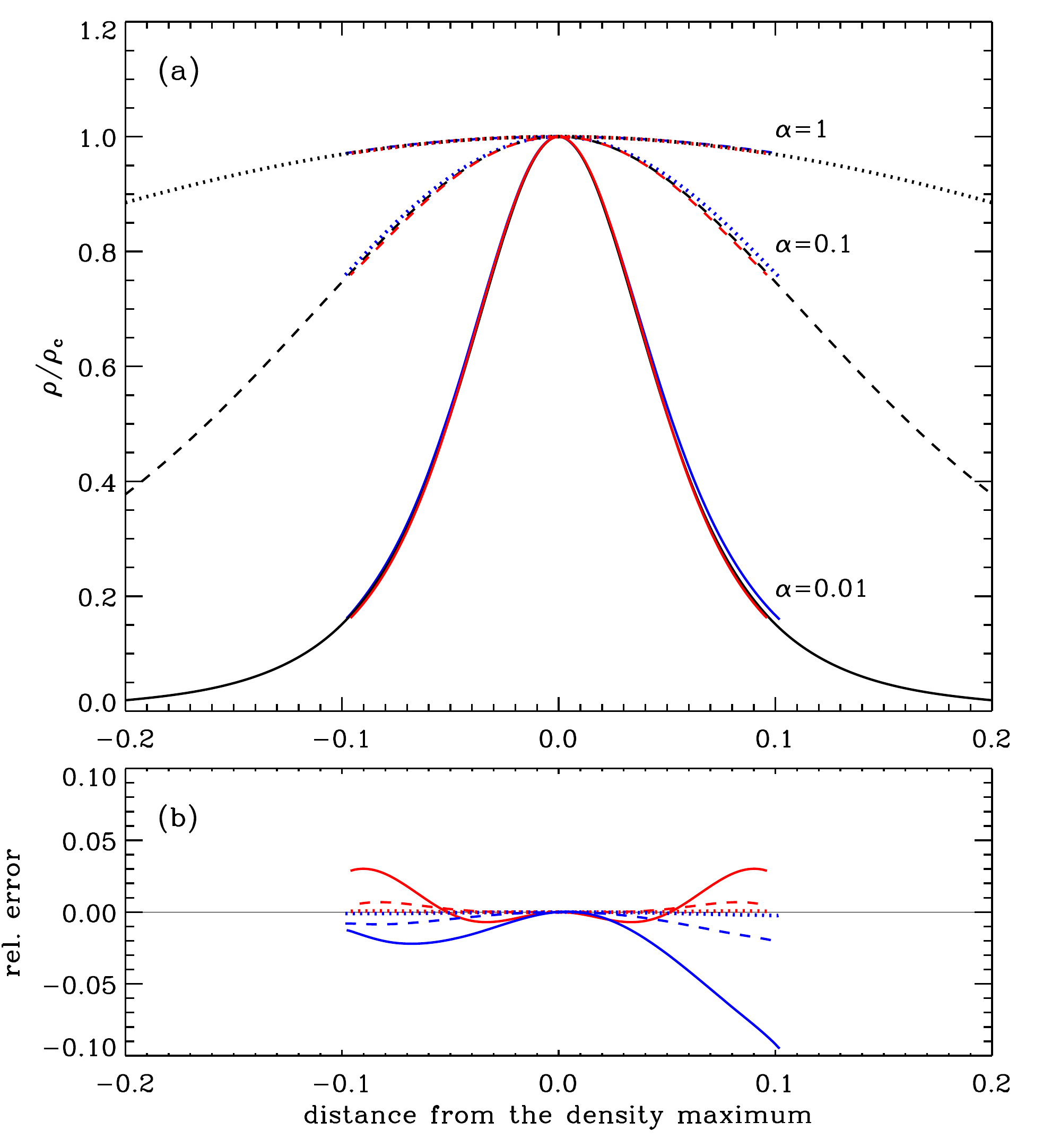}
 \caption{
(a) Equilibrium density profiles of slender rings with $\hatRmin=-0.8$ along the radial (blue) and vertical (red) directions measured from the density maximum for $\alpha=1$ (dotted), 0.1 (dashed), and 0.01 (solid), compared to the respective approximate solutions $\rho_{\rm sr}$ (black) given by Equation \eqref{e:rhosr}. (b) Relative errors, $\rhosr/\rho-1$, of the approximate solutions to the SCF results.
 \label{f:slprofile}}
\end{figure}

As the bottom panels of Figure \ref{f:isocontour} show, the density
distributions of slender rings at the meridional plane appear almost
circularly symmetric with respect to the point $(R,z)=(R_0, 0)$.
\citet{ost64b} showed that the meridional density distribution is, to
the zeroth order in $\eta_{1/2}$, given by
 \be\label{e:rhosr}
   \rhosr = \frac{\rhomax}{[1+ (\eta/H)^2/8]^2},
 \ee
where $\eta$ denotes the distance from the density maximum and
 \be
   H\equiv {\cs}/{(4\pi G \rhomax)^{1/2}}
 \ee
is the characteristic ring thickness. Note that Equation
\eqref{e:rhosr} is also the solution for non-rotating isothermal
cylinders of infinite length along its symmetry axis (e.g.,
\citealt{ost64a,nag87, inu92}). Figure \ref{f:slprofile}(a) compares
Equation \eqref{e:rhosr} (black) with the density profiles from the SCF method for slender rings with $\hatRmin=-0.8$ (or with $\hateta_0=0.1$) along the radial (blue) and vertical (red) directions from the density maximum.  The cases with $\alpha=1$, 0.1, and 0.01 are shown as dashed, dotted, and solid lines, respectively.  The relative errors, $\rhosr/\rho-1$, given in Figure \ref{f:slprofile}(b) are only a few percents in the most of the dense regions, demonstrating that Equation \eqref{e:rhosr} is a good approximation to the true density distributions of slender isothermal rings.

\section{Gravitational Instability of Slender Rings}\label{s:GI}

We now analyze gravitational instability of an isothermal ring with
$\eta_0/R_0\ll1$. As a background density distribution, we take
 \be\label{e:den0}
    \rho_0 =
     \begin{cases}
     \rhosr  ,\;\; &\textrm{for} \;\eta \leq \eta_0, \\
      0,   \;\; &\textrm{otherwise}.
  \end{cases}
  \ee
The ring is rotating at angular frequency of $\Omega_0$ mostly due
to the external gravity, such that the initial velocity is given by
${\bf v}_0 = R\Omega_0 {\bf e}_\phi$, where ${\bf e}_\phi$ is the unit vector in the $\phi$-direction.

\subsection{Perturbation Equations}\label{e:ptbeq}

The basic hydrodynamic equations governing evolution of isothermal gas read
 \be\label{e:con}
   \frac{\partial \rho}{\partial t} + \nabla\cdot (\rho {\bf v}) =0,
 \ee
  \be\label{e:mom}
   \frac{\partial {\bf v}}{\partial t} + {\bf v}\cdot \nabla {\bf v} =
   -\frac{\cs^2}{\rho} \nabla\rho  - \Omega_e^2 {\bf R} - \nabla \Phi_s,
 \ee
together with Equation \eqref{e:pos0}.

\begin{figure}
\includegraphics[angle=0, width=8.5cm]{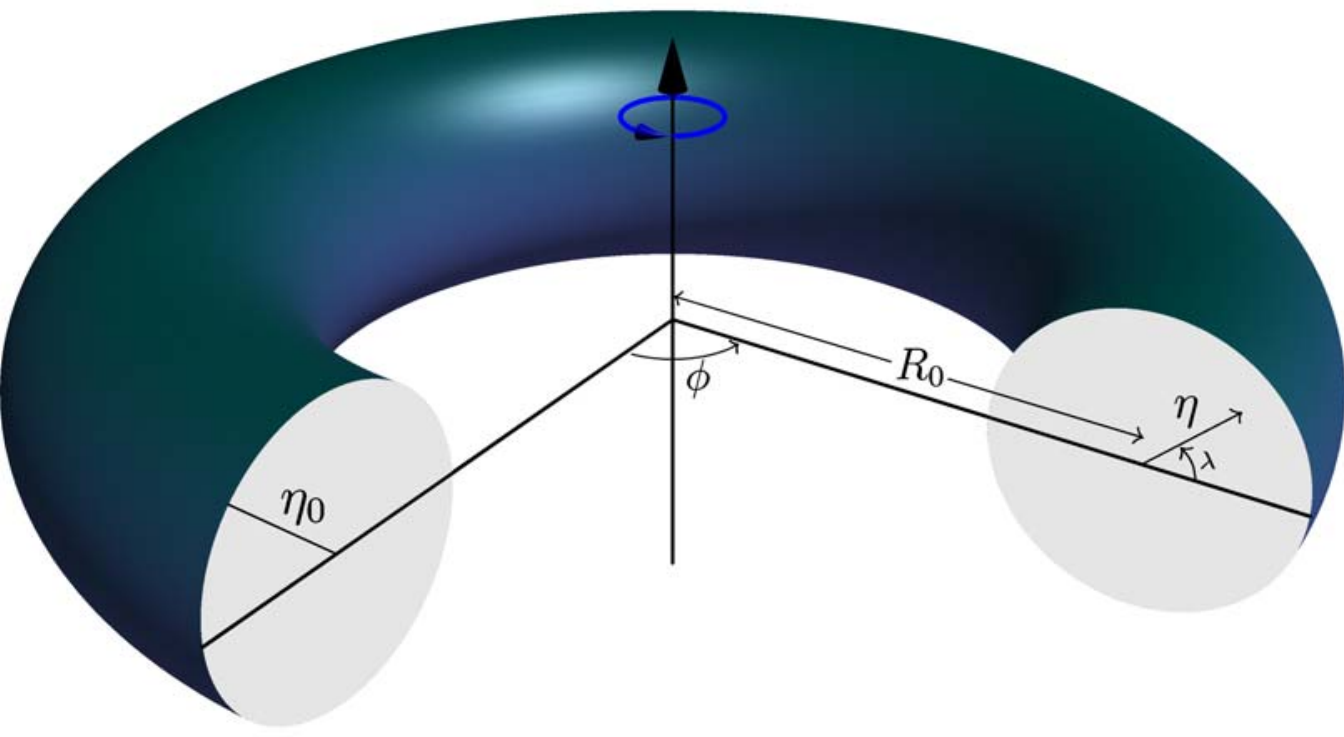}
\caption{Schematic geometry of a ring with the major axis $R_0$ and the minor axis $\eta_0$.  The coordinates $\eta$, $\lambda$, and $\phi$ measure the distance from the reference circle of radius $R_0$, the polar angle measured from the horizontal plane, and the usual cylindrical azimuthal angle, respectively.\label{f:coord}}
\end{figure}

To analyze a linear stability of a ring, it is convenient to introduce the new curvilinear coordinates $(\eta, \lambda, \phi)$, as depicted in Figure \ref{f:coord}. The new coordinates are related to the Cartesian coordinate system $(x, y, z)$ through
  \be\label{e:coord}
    \left(
  \begin{array}{c}
     x \\ y \\ z
  \end{array} \right)
   = \left(
  \begin{array}{c}
    (R_0 + \eta \cos\lambda) \cos\phi   \\
    (R_0 + \eta \cos\lambda) \sin\phi   \\
    \eta \sin\lambda
  \end{array} \right).
 \ee
The coordinate $\eta$ is the distance from a reference circle of radius $R_0$ located in the horizontal plane, while $\lambda$ is the polar angle measured from the horizontal plane; $\phi$ is the usual
cylindrical azimuthal angle. The new coordinates are orthogonal, and
reduce to the usual spherical polar coordinates $(\eta, \pi/2- \lambda, \phi)$ in the limit of $R_0/\eta \ll 1$.

Appendix \ref{a:eqn} derives gas dynamical equations for slender rings in the new curvilinear coordinates. Fluid variables are in general three-dimensional, and finding dispersion relations of
three-dimensional perturbations applied to a ring is a daunting task. However, gravitationally-unstable modes we seek in the present work are dominated by the velocity components in the $\eta$- and
$\phi$-directions, without much involving gas motions in the
$\lambda$-direction. For simplicity, therefore, we take $v_\lambda=0$
and assume that all physical quantities are independent of $\lambda$.
We also take $\cos\lambda=1$, which allows us to fully capture the
rotational effect of the ring material around the symmetry axis.
We note however that our method, by construction, is unable to handle shear that may exist in the rotation of the ring.  In Section \ref{s:sim}, we will use direct numerical simulations to show that shear does not significantly affect gravitational instabilities of slender rings.

Under these circumstances, Equations \eqref{e:con2}--\eqref{e:pos2} can be simplified to
 \be\label{e:con3}
   \frac{\partial \rho}{\partial t} + \frac{1}{\eta}\frac{\partial (\eta \rho v_\eta)}{\partial \eta}
    + \frac{1}{R_0} \frac{\partial (\rho v_\phi) }{\partial \phi} =0,
 \ee
 \be\label{e:meta3}
   \frac{\partial v_\eta}{\partial t} +
   \left( v_\eta \frac{\partial}{\partial \eta} + \frac{v_\phi}{R_0}\frac{\partial}{\partial \phi} \right) v_\eta
    - \frac{v_\phi^2}{R_0}
     = -\frac{\cs^2}{\rho}\frac{\partial \rho}{\partial \eta} - \Omega_e^2 R_0 -  \frac{\partial \Phi_s}{\partial \eta}
 \ee
 \be\label{e:mphi3}
  \frac{\partial v_\phi}{\partial t} +
  \left( v_\eta \frac{\partial}{\partial \eta} + \frac{v_\phi}{R_0}\frac{\partial}{\partial \phi} \right) v_\phi
   + \frac{v_\eta v_\phi}{R_0}
    = -\frac{\cs^2}{R_0\rho}\frac{\partial \rho}{\partial \phi}  -
    \frac{1}{R_0} \frac{\partial \Phi_s}{\partial \phi}
 \ee
 \be\label{e:pos3}
  \frac{\partial^2\Phi_s}{\partial \eta^2}
   + \frac{1}{\eta} \frac{\partial\Phi_s}{\partial \eta}
     + \frac{1}{R_0^2} \frac{\partial^2\Phi_s}{\partial \phi^2} = 4\pi G\rho.
 \ee
The initial density distribution (i.e., Eq.\ [\ref{e:den0}]) satisfies Equations \eqref{e:meta3} and \eqref{e:pos3} as long as
$\Omega_s^2 \ll \Omega_0^2$, a condition easily met for nuclear rings.

We now consider small-amplitude perturbations applied to the initial equilibrium. Denoting the background quantities and the perturbations using the subscripts ``0'' and ``1'', respectively, we linearize Equations \eqref{e:con3}--\eqref{e:pos3} to obtain
 \be\label{e:lcon}
   \left( \frac{\partial }{\partial t} +  \Omega_0 \frac{\partial}{\partial \phi} \right) \rho_1
   + \frac{1}{\eta}\frac{\partial (\eta \rho_0 v_{\eta1})}{\partial \eta}
    + \frac{\rho_0}{R_0} \frac{\partial v_{\phi1} }{\partial \phi} =0,
 \ee
 \be\label{e:lmeta}
 \left( \frac{\partial }{\partial t} +  \Omega_0 \frac{\partial}{\partial \phi} \right) v_{\eta1}
     -2 \Omega_0 v_{\phi1}
     = -\frac{\partial\chi_1 }{\partial \eta},
 \ee
 \be\label{e:lmphi}
  \left( \frac{\partial }{\partial t} +  \Omega_0 \frac{\partial}{\partial \phi} \right) v_{\phi1}
   + 2 \Omega_0 v_{\eta1}
    = -\frac{1}{R_0}\frac{\partial\chi_1 }{\partial \phi},
 \ee
 \be\label{e:lpos}
  \frac{\partial^2\Phi_{s1}}{\partial \eta^2}
   + \frac{1}{\eta} \frac{\partial\Phi_{s1}}{\partial \eta}
     + \frac{1}{R_0^2} \frac{\partial^2\Phi_{s1}}{\partial \phi^2} = 4\pi
     G\rho_1,
 \ee
where
 \be
    \chi_1 \equiv \cs^2\frac{\rho_1}{\rho_0} + \Phi_{s1}.
  \ee

We assume that any perturbation, $q_1$, varies in space and time as
  \be\label{e:ptb}
   q_1 (\eta, \phi, t) = q_1(\eta) \exp( im\phi - i\omega t ),
 \ee
with $\omega$ and $m$ being the frequency and azimuthal mode number of the perturbations, respectively. Substituting Equation \eqref{e:ptb} into Equations \eqref{e:lcon}--\eqref{e:lpos},
one obtains
 \be\label{e:lcon1}
   \frac{d v_{\eta1}}{d\eta} + \frac{d\ln(\eta\rho_0)}{d\eta} v_{\eta1}
   = i \tomega \frac{\rho_1}{\rho_0} - i\frac{m}{R_0} v_{\phi1},
 \ee
 \be\label{e:lmeta1}
    v_{\eta1} = \frac{i}{\tomega^2 - 4\Omega_0^2}
    \left(\frac{ 2m\Omega_0}{R_0}\chi_1 -\tomega \frac{d\chi_1}{d\eta}  \right),
 \ee
 \be\label{e:lmphi1}
    v_{\phi1} = \frac{1}{\tomega^2 - 4\Omega_0^2}
    \left(\frac{m\tomega}{R_0}\chi_1 - 2\Omega_0 \frac{d\chi_1}{d\eta}  \right),
 \ee
 \be\label{e:lpos1}
   \frac{d^2\Phi_{s1}}{d \eta^2}
   + \frac{1}{\eta} \frac{d\Phi_{s1}}{d \eta}
   - \frac{m^2}{R_0^2} \Phi_{s1} = 4\pi G\rho_1,
 \ee
where $\tomega \equiv \omega - m\Omega_0$ is the Doppler-shifted
frequency. Eliminating $v_{\eta1}$ and $v_{\phi1}$ in favor of $\chi_1$ from Equations \eqref{e:lcon1}--\eqref{e:lmphi1}, one finally obtains
  \be\label{e:lcon2}
  \begin{split}
   \frac{d^2\chi_1}{d\eta^2} +
   \frac{d\ln(\eta\rho_0)}{d\eta} \frac{d\chi_1}{d\eta} -
   \left[ \frac{2\Omega_0}{\tomega} \frac{m}{R_0} \frac{d\ln(\eta\rho_0)}{d\eta}
   + \frac{m^2}{R_0^2} \right] \chi_1  \\
         = -(\tomega^2-4\Omega_0^2) \frac{\rho_1}{\rho_0}.
   \end{split}
  \ee
Equations \eqref{e:lpos1} and \eqref{e:lcon2} constitute a set of
coupled equations that can be solved simultaneously for eigenvalue
$\omega$ subject to proper boundary conditions.

\subsection{Boundary Conditions}\label{e:bdcon}

Since Equations \eqref{e:lpos1} and \eqref{e:lcon2} are second-order
differential equations, we need to have five constraints in order to
determine $\omega$ unambiguously. Since this is a linear problem, we
are free to choose the amplitude of one variable arbitrarily. Two
conditions come from the inner boundary by the requirements
  \be\label{e:bd1}
    \left.\frac{d\chi_1}{d\eta}\right|_{\eta=0}  =
    \left.\frac{d\Phi_{s1}}{d\eta}\right|_{\eta=0} =0,
  \ee
for regular solutions at $\eta=0$.

The remaining two conditions can be obtained from the outer boundary.
Perturbations given in Equation \eqref{e:ptb} also disturb the ring
surface to
  \be\label{e:bdsurf}
     \eta = \eta_0 + \eta_1 \exp(im\phi - i\omega t ),
  \ee
with amplitude $\eta_1$, implying that
  \be\label{e:velbd}
     v_{\eta1} (\eta_0) = -i\omega \eta_1.
  \ee
The pressure equilibrium at the disturbed surface, $\cs^2
\rho_0(\eta_0)+P_1(\eta_0)=P_{\rm ext}$, with $P_1$ representing the
perturbed pressure (e.g., \citealt{nag87}), requires
  \be\label{e:3rdbd}
      \left.\frac{d\rho_0}{d\eta}\right|_{\eta_0} \eta_1 + \rho_1(\eta_0) = 0,
  \ee
which is a third boundary condition.

To derive a fourth boundary condition, we follow \citet{gol65} to
assume that the perturbed mass near the outer boundary is restricted to a thin annulus such that $\rho_1 =\rho_0\eta_1\delta (\eta-\eta_0)$ (see also, \citealt{nag87,kimjg}). Integrating Equation \eqref{e:lpos1} from $\eta=\eta_0$ to $\eta=\eta_0 + \eta_1$, one obtains
  \be\label{e:pbd}
    \frac{d\Phi_{s1}^+}{d\eta}   -
    \frac{d\Phi_{s1}^-}{d\eta}  = 4\pi G\rho_0\eta_1\;\;\;\text{at}\;\;\;
    \eta=\eta_0,
  \ee
where the superscripts ``$+$'' and ``$-$'' indicate the potentials evaluated just outside and inside the ring surface, respectively. Assuming that the region outside the ring is filled with an extremely hot, tenuous gas, $\Phi_{s1}^+$ should satisfy Equation \eqref{e:lpos} with $\rho_1=0$. The regular solution at infinity can be expressed as
  \be\label{e:extp}
    \Phi_{s1}^+ = A K_0 \left( \frac{m}{R_0} \eta \right),\;\;\;\text{for}\;\;\;\eta/\eta_0 \geq 1,
  \ee
where $A$ is a constant and $K_n$ is the second-kind modified Bessel
function of order $n$. The condition that the gravitational potential
should be continuous across the surface gives $A=
\Phi_{s1}(\eta_0)/K_0(m\eta_0/R_0)$. Plugging Equation \eqref{e:extp}
into Equation \eqref{e:pbd} gives
  \be\label{e:4thbd}
     \frac{d\Phi_{s1}}{d\eta} + 4\pi G\rho_0\eta_1 = -\frac{m}{R_0}\frac{K_1(m\eta_0/R_0)}{K_0(m\eta_0/R_0)} \Phi_{s1}
     \;\;\;\text{at}\;\;\; \eta=\eta_0.
  \ee
Equations \eqref{e:bd1}, \eqref{e:3rdbd}, and \eqref{e:4thbd} are our
complete set of the boundary conditions.

\subsection{Method of Computation}\label{e:mtd}

By writing Equations \eqref{e:lpos1} and \eqref{e:lcon2} into a
dimensionless form, one can see that the problem of finding the
dimensionless eigenvalue $\hatomg \equiv \omega/(G\rhomax)^{1/2}$ is
completely specified by four dimensionless parameters, $\eta_0/R_0$,
$R_0/H$, $\hatOmega \equiv \Omega_0/(G\rhomax)^{1/2}$, and $m$. Nuclear rings typically have $\eta_0\sim 0.1\kpc$, $R_0\sim1\kpc$,
$M\sim4\times 10^8\Msun$, and $\Omega_0\sim 100$--$200
\kms\;\kpc^{-1}$. The corresponding mean density is $\hatavgrho =
M/(2\pi^2\eta_0^2R_0) \sim 1\times 10^{-22} \rm\;g\;cm^{-3}$, and the
characteristic  ring thickness is
 \be
   H = 35 \pc\;\left(\frac{\cs}{10 \kms}\right)
        \left( \frac{\rhomax}{1\times 10^{-22} \rm\;g\;cm^{-3}} \right)^{-1/2},
 \ee
We therefore choose $\eta_0/R_0=0.1$ and $R_0/H=30$ as our standard set of parameters, but also vary $R_0/H$ and $\hatOmega$ to explore the effects of ring thickness and rotation on the ring stability.  For slender rings with $\eta_0/R_0=0.1$,  $R_0/H$ is related to $\alpha$ through $\alpha \simeq 10.4 \left({R_0}/{H}\right) ^{-2}$.

\begin{figure*}
\includegraphics[angle=0, width=18cm]{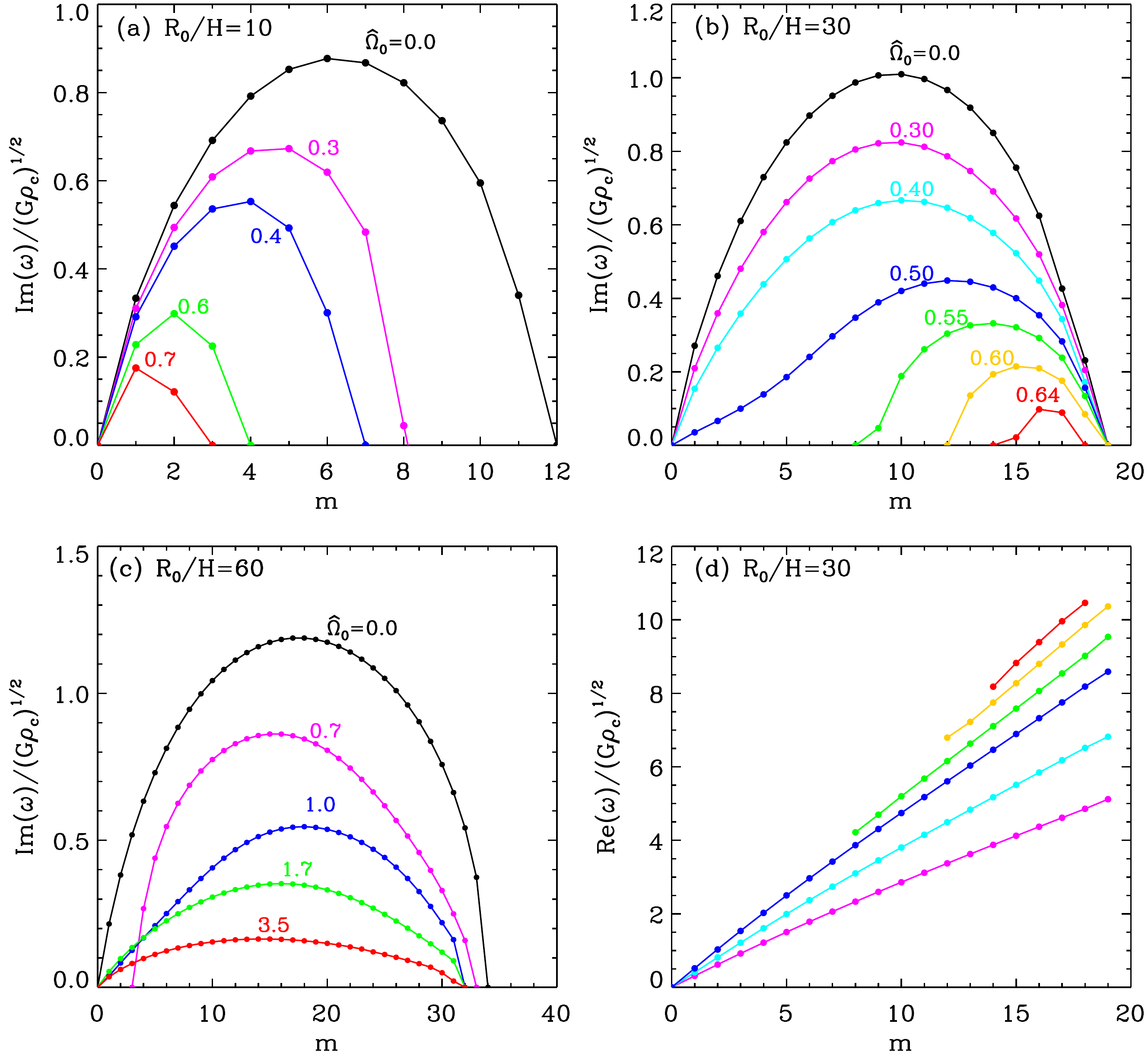}
\caption{(a)--(c)
Imaginary parts of the eigenfrequencies of the unstable modes for
various values of the angular frequency $\hatOmega$ in the rings with
$R_0/H=10$, 30, and 60, and (d) the real parts of the unstable
eigenfrequencies for the rings with $R_0/H=30$. For all cases, the
ratio of the ring minor to major axes is fixed to $\eta_0/R_0=0.1$. A
ring with larger $R_0/H$ (or smaller $\alpha$) is more unstable, and
thus has a larger growth rate and a larger unstable range of $m$.
Rotation tends to reduce the growth rate.
 \label{f:rdisp_all}}
\end{figure*}

As a normalization condition, we take ${\rm Re}(\chi_1)={\rm
Im}(\chi_1)=1$ at the outer boundary. For given $m$ and $\hatOmega$, we first choose two trial values  for $\hatomg$ and $\Phi_{s1}$ at
$\eta=\eta_0$, and calculate $d\chi_1/d\eta$ and $d\Phi_{s1}/d\eta$ at the outer boundary from Equations \eqref{e:3rdbd} and \eqref{e:4thbd}. Next, we integrate Equations \eqref{e:lpos1} and \eqref{e:lcon2} from $\eta=\eta_0$ to $\eta=0$ and check if the two conditions in Equation \eqref{e:bd1} are satisfied. If not, we vary $\Phi_{s1}(\eta_0)$ and $\hatomg$ one by one and repeat the calculations until the inner boundary conditions are fulfilled within tolerance.

\subsection{Dispersion Relations}\label{s:result}

Figure \ref{f:rdisp_all}(a)--(c) plots the imaginary parts of
eigenfrequencies for gravitationally unstable modes for isothermal
rings with $R_0/H=10$, 30, and 60 (or $\alpha=1.04\times 10^{-1}$,
$1.15\times10^{-2}$, and $2.88\times10^{-3}$) as functions of the
azimuthal mode number $m$ and the rotational angular frequency
$\hatOmega$. For all cases, the ring thickness is fixed to
$\eta_0/R_0=0.1$. When $\hatOmega=0$, the maximum growth rates are
Im$(\hatomgmax)=0.88$, 1.01, and 1.19, which are achieved at $m_{\rm
max}=6$, 10, and 18, for the rings with $R_0/H=10$, 30, and 60,
respectively. Note that the dispersion relations with $\hatOmega=0$ are identical to those of axisymmetric modes in an infinite isothermal cylinder, presented by \citet{nag87}, as long as $m/R_0$ is replaced with the vertical wavenumber $k$. Without rotation, $\hatomg$ is always an imaginary number, corresponding to pure instability. When $\hatOmega\neq 0$, however, eigenfrequencies are complex numbers with the real parts almost linearly proportional to $m$, corresponding to overstability, as exemplified in Figure \ref{f:rdisp_all}(d) for $R_0/H=30$. This is a generic property of any instability occurring in a non-static background medium (e.g., \citealt{mat94,kim14a,kim15}).

\begin{figure}
\includegraphics[angle=0, width=8.5cm]{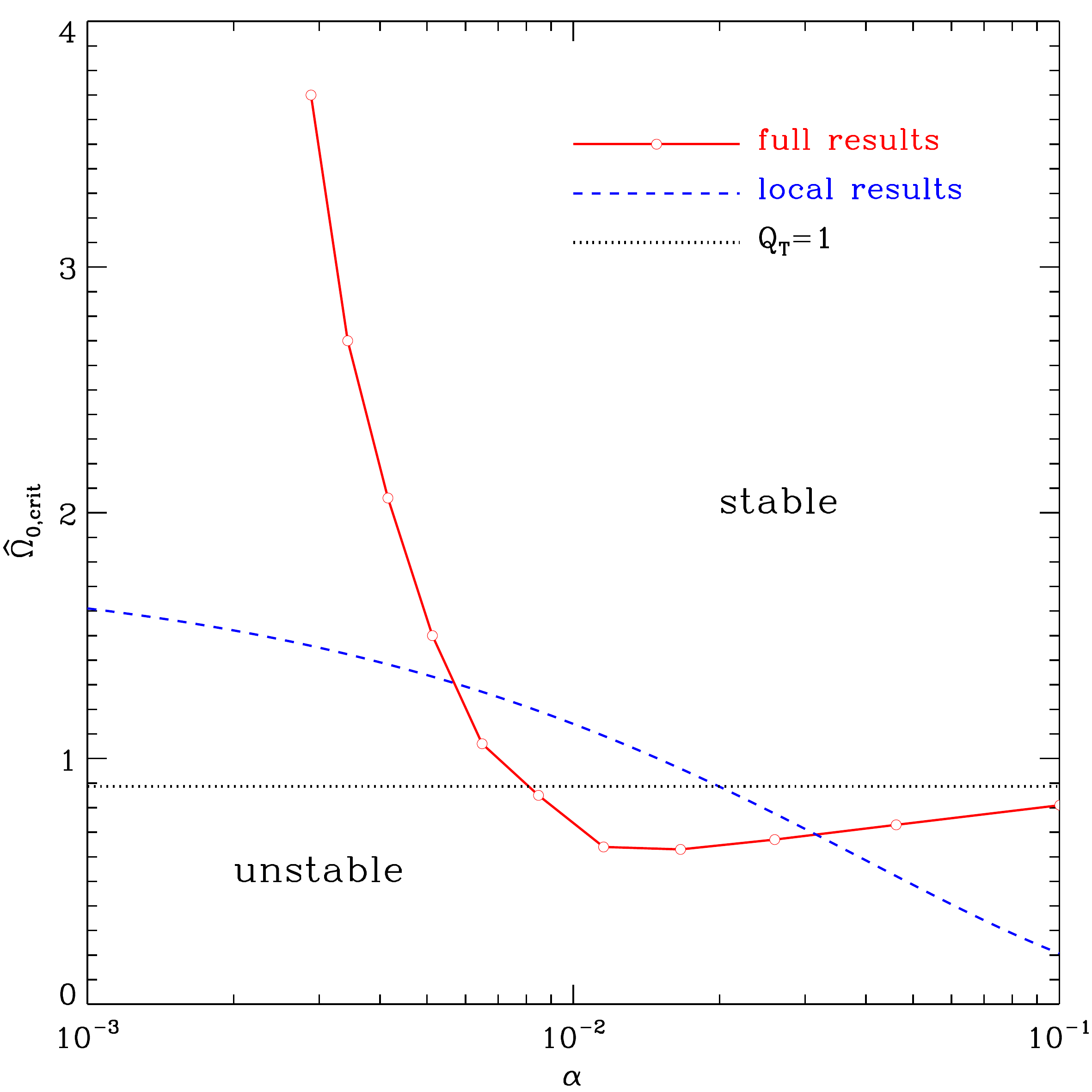}
 \caption{Critical angular
frequencies  as a function of $\alpha$, with the
upper-right region corresponding to stable configurations.
The red solid line with dots is the results of our full stability analysis, while the blue dashed line draws Equation \eqref{e:lcrit} from the local dispersion relation. The critical frequency from the Toomre condition is given as a horizontal dotted line for comparison. See text for details.
 \label{f:crit_cal}}
\end{figure}

It is apparent that a ring with larger $R_0/H$ (or smaller $\alpha$) is more unstable owing to a smaller sound speed and/or a larger ring mass. Overall, rotation tends to stabilize the instability, reducing both the growth rate and the unstable range of $m$, although their dependence on $\hatOmega$ is not simple. When $R_0/H=10$ (or 30), the reduction of the growth rate due to rotation is larger at larger (or smaller) $m$, making $m_{\rm max}$ shifted to a smaller (or larger) value as $\hatOmega$ increases. In the case of $R_0/H=60$, on the other hand, rotation simply reduces the growth rate, without much affecting the unstable range of $m$. The instability is completely suppressed when $\hatOmega\gtrsim 0.81$, 0.64, and 3.70 for $R_0/H=10$, 30, and 60, respectively.
Figure \ref{f:crit_cal} plots the critical angular frequency $\hatOmegac$ against $\alpha$ as the solid line, with the upper-right region corresponding to the stable regime. Note that $\hatOmegac$ is almost constant at $\sim 0.7$ for $\alpha\gtrsim0.01$ and increases rapidly as $\alpha$ decreases.

It is interesting to compare our results for $\hatOmegac$ with those from other simple estimates. The Toomre stability parameter $Q_T=\kappa_0c_s/(\pi G \Sigma_0)$ has usually been invoked to judge whether a flattened system under consideration is gravitationally stable or not.  For a ring, it is unclear how to choose $\Sigma_0$ since the ring surface density varies with $R$.  If we simply take $\Sigma_0=2\rho_c H$, about a half of the maximum surface density $\Sigma_{\rm max}=2\int_0^\infty\rhosr d\eta = 2^{1/2}\pi \rho_c H$ across the ring center, the Toomre condition for marginal stability $Q_T=1$ corresponds to $\hatOmegac = (\pi/4)^{1/2}\approx 0.89$, independent of $\alpha$, which is plotted as the horizontal dotted line in Figure \ref{f:crit_cal}.  The critical angular frequency from the Toomre condition is close to the results of our stability analysis for $\alpha \gtrsim 0.01$, but deviates considerably for smaller $\alpha$. Strictly speaking, the Toomre condition is valid only for thin disks that are infinitesimally thin in the vertical direction but infinitely extended in the horizontal direction. Even the thin-disk gravity underestimates self-gravity of highly concentrated rings with $\alpha\lesssim 0.01$.

\citet{elm94} presented a local dispersion relation for gravitational instability of nuclear rings by treating them as being thin and locally cylindrical. In Appendix \ref{a:wkb}, we solve Equations \eqref{e:lpos1} and \eqref{e:lcon2} for local waves that vary very rapidly in the azimuthal direction (i.e., $m/R_0 \gg |d\ln(\eta\rho_0)/d\eta|$) but remain constant in the $\eta$-direction (i.e., $d\chi_1/d\eta=0$). The resulting dispersion relation (Eq.~\eqref{e:ldisp}) is the same as the one given in \citet{elm94} (in the absence of magnetic fields and gas accretion).  The critical angular frequency for local waves is then given by
  \be\label{e:lcrit}
    \hatOmegac^2 =  \max_m \left\{  \pi \left[ 1 -  \frac{m\eta_0}{R_0} K_1\left( \frac{m\eta_0}{R_0} \right) \right] -  \frac{\alpha}{4} m^2 \right\},
  \ee
which is plotted in Figure \ref{f:crit_cal} as the blue dashed line for $\eta_0/R_0=0.1$. Although $\hatOmegac$ from Equation \eqref{e:lcrit} is similar to the results of the full analysis for $\alpha\sim 3\times 10^{-2}$, it underestimates the latter considerably for $\alpha \lesssim 5\times 10^{-3}$.
This is because rings with smaller $\alpha$ are increasingly more strongly concentrated that the approximations of constant $\rho_0$ and $\chi_1$ over $\eta$ become invalid.\footnote{The critical density $\rho_{\rm crit}=0.6\kappa^2/G$ of \citet{elm94}, or equivalently $\hatOmegac= 0.65$ in our notation,  mentioned in Section \ref{s:intro} was based on the assumption of $m \eta_0/R_0 \lesssim 1$, which conflicts with the local approximation employed in the derivation of Equation \eqref{e:lcrit}. In view of Figure \ref{f:rdisp_all}, this  cannot capture the most unstable modes for small $\alpha$, as well.}  For $\alpha \propto (H/R_0)^2 \rightarrow 0$, rings can be approximated as strongly concentrated cylinders for which motions along the symmetry axis do not affect their gravitational instability much. Very large values of $\hatOmega$ are required to stabilize such small-$\alpha$ rings.

We thus conclude that both the Toomre condition and the local results
cannot adequately describe the critical angular frequencies of nuclear rings, especially when $\alpha$ is very small. In Section \ref{s:dis}, we will apply the stability curve derived from the full stability analysis to observed rings in real galaxies.

\begin{figure*}
\hspace{1cm}
\includegraphics[angle=0, width=16cm]{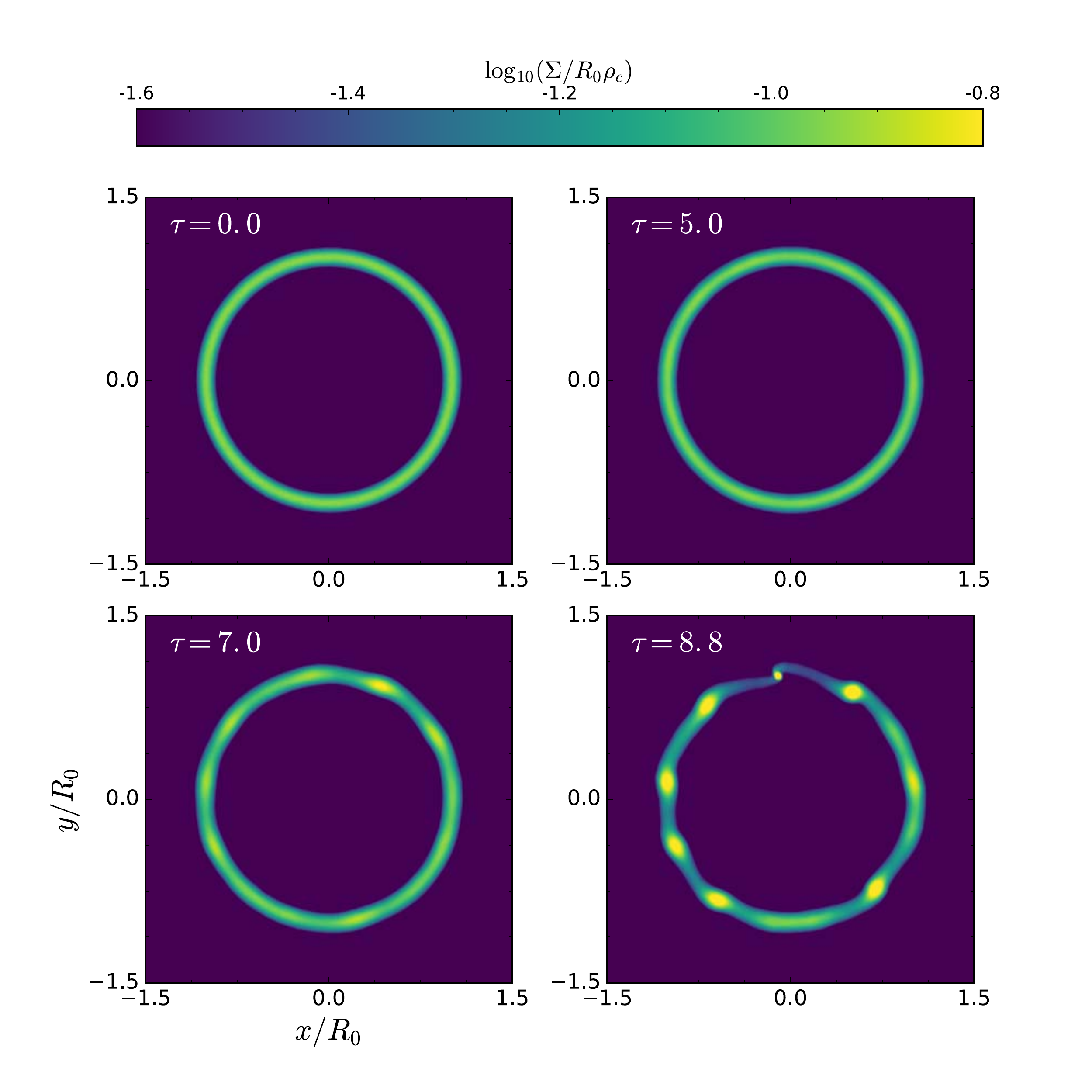}
\vspace{-0.2cm}
\caption{Snapshots of the projected density $\Sigma=\int\rho dz$ on to the equatorial plane of  the
rigidly-rotating ($q=0$) model with $R_0/H=30$, $\eta_0/R_0=0.1$, and $\hatOmega=0.30$ at $\tau=0.0$, 5.0, 7.0, and 8.8. As a result of gravitational instability, the ring fragments into 11 clumps.
 \label{f:proj}}
\end{figure*}

\begin{figure}
\includegraphics[angle=0, width=8.5cm]{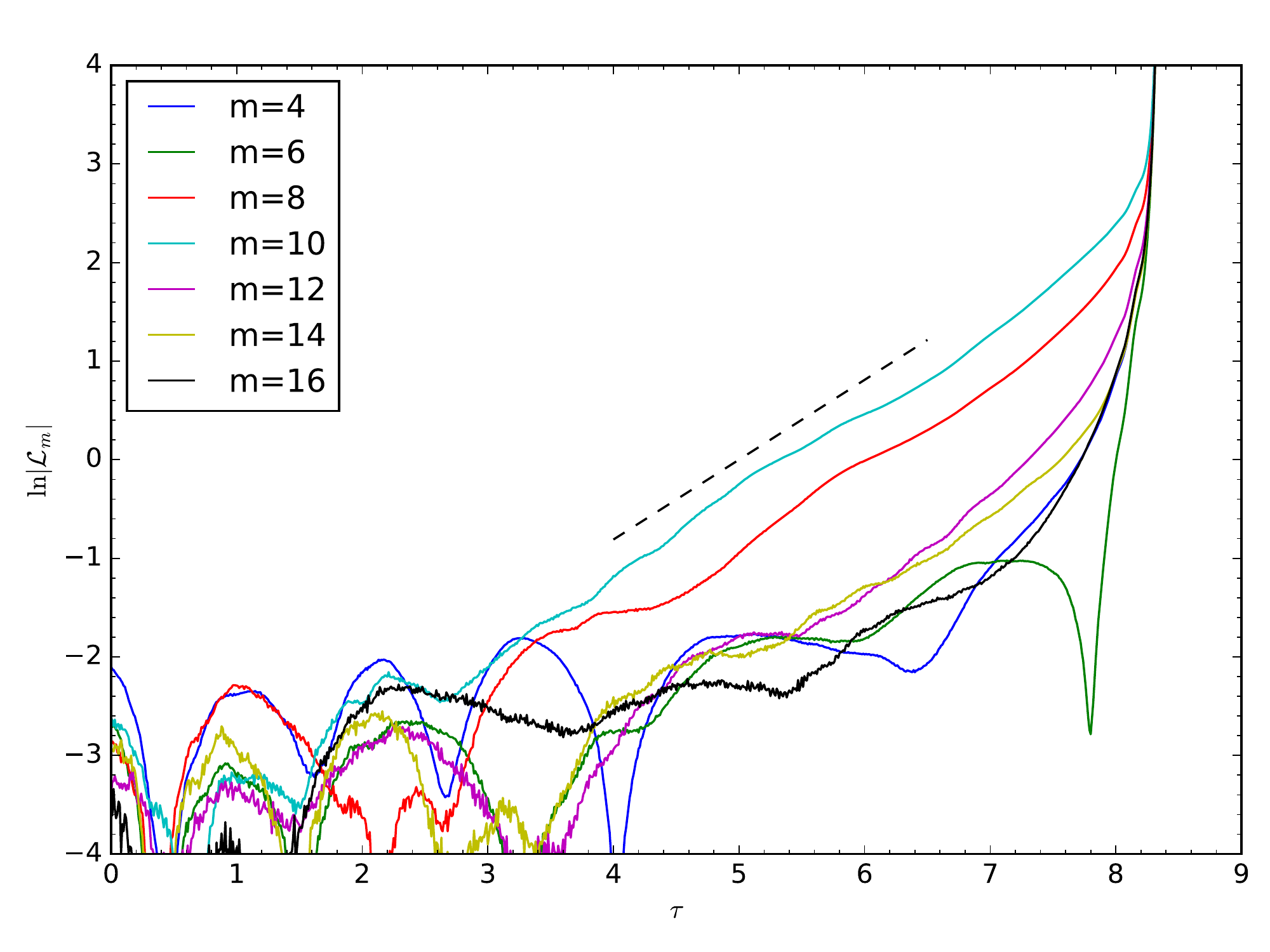}
\caption{Temporal evolution of the amplitudes of even-$m$ Fourier modes for the $q=0$ model shown in Figure \ref{f:proj}. The dashed-line segment indicates a slope of $0.81$, very close to the growth of $m=8$--12 modes, consistent with the results of the linear stability analysis.
 \label{f:evol}}
\end{figure}

\subsection{Nonlinear Simulations}\label{s:sim}

To check the results of our linear stability analysis as well as to explore the effect of differential rotation, we conduct direct numerical simulations for gravitational instability of a slender isothermal ring using the GIZMO code \citep{hop15,hop16}. GIZMO is a second-order accurate magnetohydrodynamics code based on a Lagrangian, mesh-free, Godunov-type method, and conserves mass, momentum, and energy almost exactly.  In our calculations, the basic
equations of hydrodynamics are solved by the Meshless Finite-Mass Method known to conserve angular momentum very well.

To obtain an initial particle distribution, we first use the SCF method to construct the equilibrium density distribution of the rigidly-rotating ring with $R_0/H=30$ and $\eta_0/R_0=0.1$ in the absence of external gravity. The rotation frequency of this ring is found to be $\hatOmegas=0.22$. The initial particle positions are then sampled by a rejection technique that uses Halton's quasi-random sequences over usual random numbers in order to reduce Poisson noises (e.g., \citealt{pre88}).  We then impose a radially-variying external gravity $\hatOmegae^2 (R) {\mathbf{\widehat R}}$ to boost the angular velocity of the ring particles according to
  \be\label{e:diffOmg}
   \hatOmega=(\hatOmegas^2+\hatOmegae^2)^{1/2}= 0.30 \left(\frac{R}{R_0}\right)^{-q}.
  \ee
Here, $q$ is the rate of shear in the ring rotation, such that
$q=0$ and $1$ corresponds to rigid-body and flat rotation, respectively. We vary $q$ from 0 to 1.5 to study the effect of shear on the growth of gravitational instability.

\begin{figure*}
\hspace{1.5cm}\includegraphics[angle=0, width=15cm]{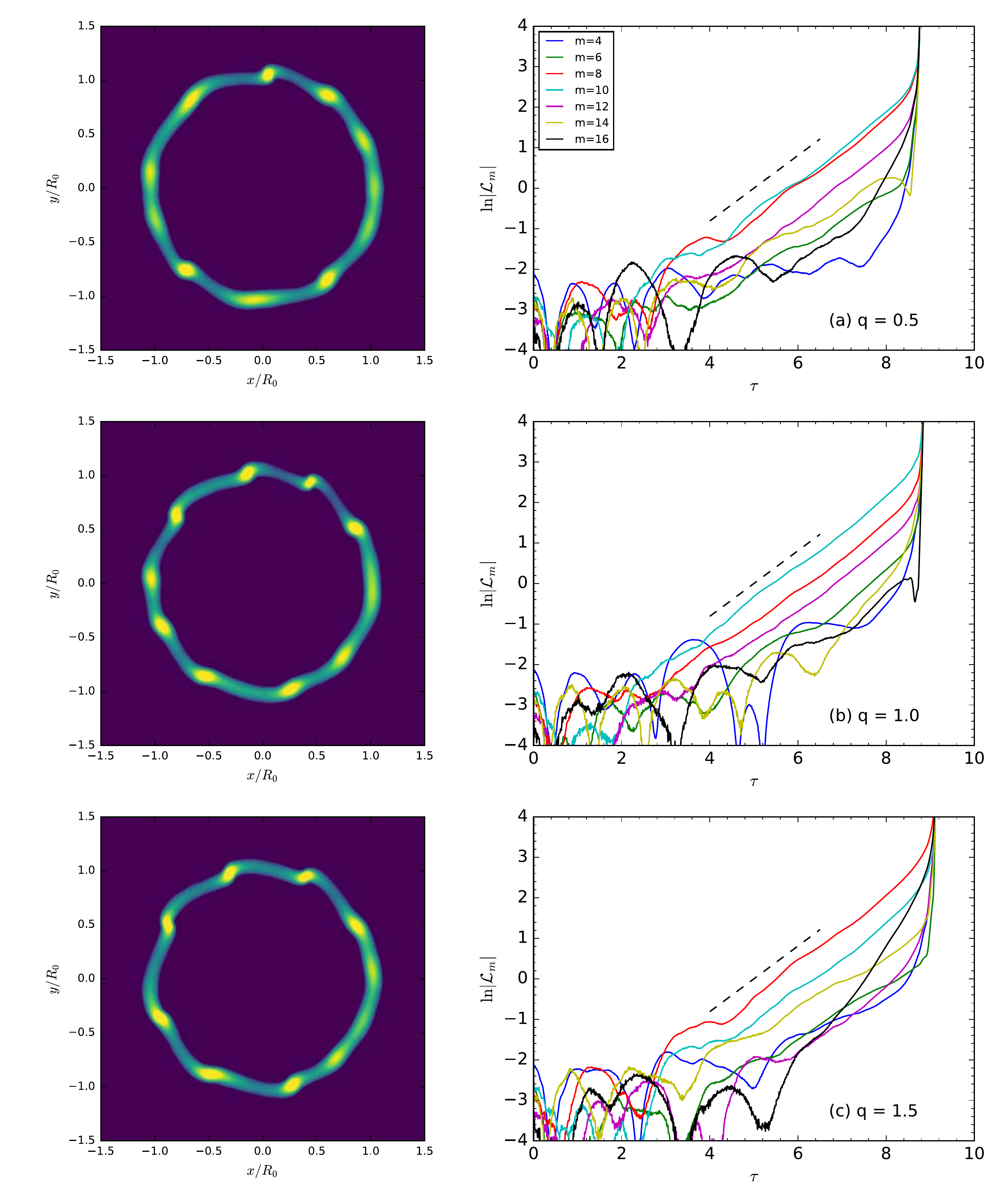}
\caption{ Snapshots of the projected density (left)
at $\tau=8.8$, $9.0$, and $9.1$, and the temporal variations of the amplitudes of even-$m$ Fourier modes (right) in models with $q=0.5$, $1.0$, and $1.5$ from top to bottom.  The dashed-line segment in each panel indicates a slope of $0.81$, which describes the growth rates of dominant modes fairly well for all models.  The number of clumps produced is 11, 10, and 10 from top to bottom.
 \label{f:allq}}
\end{figure*}

To represent a hot tenuous external medium, we also distribute low-mass particles/cells outside the ring but inside a cylindrical volume with radius $R/R_0 = 2$ and height $h/R_0 = 2$.
We let the external medium follow the rotation law given in Equation \eqref{e:diffOmg}, and adjust its density distribution to balance the centrifugal force and gravity of the ring.  We ensure a pressure equilibrium between the ring and the external medium that has 100 times lower density than the ring at the contact surfaces.  The number of particles/cells for the ring and external medium is $5\times10^5$ each.

At the very initial phase of evolution ($\tau\equiv t(G\rhomax)^{1/2} \lesssim 0.25$), we iron out any residual Poisson sampling noises by introducing an artificial damping force $f \propto -v_\eta \widehat \eta$. The system thus evolves from a smooth, steady equilibrium state, without undergoing violent expansion or contraction, and gradually picks up gravitationally unstable modes.
Figure \ref{f:proj} plots snapshots of the projected density $\Sigma=\int\rho dz$ onto the equatorial plane at $\tau=0$, 5, 7, and 8.8 for a rigidly-rotating model with $q=0$.  Defining  the line density as ${\mathcal L} \equiv \int \Sigma dR$, we calculate the amplitude ${\mathcal L} _m$ of each azimuthal mode $m$ via Fourier transform at each time. Figure \ref{f:evol} displays ${\mathcal L}_m$ as functions of $\tau$ for even-$m$ modes in the $q=0$ model.  Note that the modes with $m=8-12$ dominate during the linear growth phase ($4\lesssim \tau\lesssim 6$), resulting in 11 clumps at the end of the run.  The growth rates of these modes are all similar at $\sim (0.80$--$0.82) (G\rhomax)^{1/2}$, as indicated as the dashed-line segment, consistent with the linear results shown in Figure \ref{f:rdisp_all}(b). This indirectly confirms that the assumptions made in our stability analysis are quite reasonable.

The left panels of Figure \ref{f:allq} plot the snapshots of surface density in the equatorial plane at the end of the runs (at $\tau=8.8$, $9.0$, and $9.1$) for models with $q=0.5$, $1.0$, and $1.5$ from top to bottom, while the right panels give the temporal evolution of ${\mathcal L} _m$ for even-$m$ modes. In the $q=0.5$ model, the $m=8$ and 10 modes dominate almost equally,  while the models with $q=1.0$ and 1.5 are dominated by the $m=10$ and $m=8$ mode, respectively. Note that the growth rates of the dominant modes in all models are very close to $0.81 (G\rhomax)^{1/2}$, marked by the dashed-line segment in each panel. The number of clumps produced as a result of gravitational instability is 10 or 11,
insensitive to $q$, demonstrating that shear does not affect the character of gravitational instability of slender rings.

\section{Summary \& Discussion}\label{s:sum}

\subsection{Summary}

Nuclear rings at centers of barred galaxies exhibit strong star
formation activities. They are thought to undergo gravitational
instability when sufficiently massive. To study their equilibrium
properties and stability to gravitational perturbations, we approximate nuclear rings as isothermal objects. We modify the SCF method of \citet{hac86a} to make it suitable for an isothermal equation of state, and construct equilibrium sequences of rigidly-rotating, self-gravitating, isothermal bodies. A steady equilibrium is uniquely specified by two dimensionless parameters: $\alpha$ and $\hatrB$ (see Eqs.~[\ref{e:alp}] and [\ref{e:rB}]). The former is the measure of the thermal energy relative to gravitational potential energy of an equilibrium body, while the latter corresponds to the ellipticity for spheroid-like configurations or the thickness for ring-like configurations. We take a convention that $\hatrB$ is positive (or negative) for spheroid-like (or ring-like) objects.

To test our SCF method, we first apply it to the case of rotating
incompressible bodies, and confirm that our method is able to reproduce the Maclaurin spheroid sequence when $0.158 \leq \hatrB \leq1$. With improved resolution, our method gives more accurate results than those obtained by \citet{eri81} and \citet{hac86a} for the concave hamburger sequence with $0 \lesssim \hatrB < 0.158$ .  Our method also successfully reproduces isothermal Bonnor-Ebert spheres, with larger $\alpha$ corresponding to a higher degree of central density concentration.

We then use our SCF method to obtain the density distributions of
rotating isothermal equilibria on the meridional plane, as illustrated in Figure \ref{f:isocontour}. We calculate the dependence on $\hatrB$ of various dimensionless quantities such as the rotational angular frequency $\hatOmegas$, the total mass $\hatM$, the mean density $\hatavgrho$, the total kinetic energy $\hatT$, and the gravitational potential energy $\hatW$. These values are tabulated in Table \ref{t:iso} and given graphically in Figure \ref{f:isoRB}.  We find that an equilibrium density profile is more centrally concentrated for smaller $\alpha$. Unlike the incompressible bodies, not all values of $\hatrB$ result in an isothermal equilibrium configuration. Spheroid-like equilibria exist only for $\hatrBmax \leq \hatrB \leq 1$, while ring-like (or hamburger-like) configurations are possible only for $-1< \hatrB<\hatrBmin$: otherwise, the centrifugal potential is too
large to form gravitationally bound objects. The critical $\hatrB$
values are found to be $\hatrBmax=0.27$, 0.51, and 0.59, and
$\hatrBmin=0.13$, $-0.14$, and $-0.40$ for $\alpha=1$, $0.1$, and
$0.01$, respectively.

In general, $\hatOmegas$ is a decreasing function of $|\hatrB|$. This
is naturally expected for spheroid-like configurations since faster
rotation leads to a more flattened equilibrium.  As $\hatrB$ approaches $-1$, on the other hand, ring-like configurations becomes less massive and thus requires weaker centrifugal force to balance self-gravity. Due to stronger central concentration, $\hatOmegas$, $\hatM$, $\hatavgrho$, $\hatT$, and $|\hatW|$ all become smaller as $\alpha$ decreases. For $\alpha< 0.1$, $\hatM$ and $\hatW$ are insensitive to $\hatrB\gtrsim0.6$ for spheroid-like equilibria since the density in the outer parts becomes vanishingly small. For a given value of the normalized angular momentum $j$, the normalized angular frequency $\omega_s$ becomes smaller with decreasing $\alpha$, although the energy ratio $T/|W|$ is insensitive to $\alpha$.

\begin{deluxetable*}{rcccccccc}[!t]
\tablecaption{Properties of Observed Nuclear Rings\label{t:obsring}}
\tablewidth{0pt} %
\tablehead{  & \colhead{$R_1$}
          &   & \colhead{$v_{\rm rot}$}
          & \colhead{$M_g$}
          &
          &
          &
          &  \\
      \colhead{Galaxy}    & \colhead{(kpc)}
          &  \colhead{$e$}  & \colhead{(km s$^{-1}$)}
          & \colhead{$(10^7\Msun)$}
          & \colhead{Age Grad.}
          &  \colhead{$\alpha$}
          &  \colhead{$\hatOmega$}
          &   \colhead{Ref.} \\
            \colhead{(1)}   & \colhead{(2)}
          & \colhead{(3)}   & \colhead{(4)}
          & \colhead{(5)}
          & \colhead{(6)}
          & \colhead{(7)}
          & \colhead{(8)}
          & \colhead{(9)} }
\startdata
NGC  473  &  1.69  &  0.06  & 125  &  40  &   Yes   & 1.69E$-2$   &  1.71   &      \\
NGC  613  &  0.40  &  0.26  & 115  &  40  &     ?   & 3.93E$-3$   &  0.76   &      \\
NGC 1097  &  0.97  &  0.32  & 220  & 140  &    No   & 2.70E$-3$   &  1.20   &  (a),(b)    \\
NGC 1300  &  0.40  &  0.15  & 155  &  40  &    No   & 3.98E$-3$   &  1.03   &      \\
NGC 1343  &  1.97  &  0.30  &  80  &  40  &   Yes   & 1.92E$-2$   &  1.17   &      \\
NGC 1530  &  1.20  &  0.80  & 180  &  40  &   Yes   & 9.30E$-3$   &  1.82   &      \\
NGC 2903  &  0.16  &  0.32  &  60  &  35  &     ?   & 1.78E$-3$   &  0.27   &  (c)    \\
NGC 3351  &  0.15  &  0.11  & 120  &  31  &     ?   & 1.89E$-3$   &  0.55   &  (c)    \\
NGC 4303  &  0.35  &  0.11  &  90  &  42  &    No   & 3.32E$-3$   &  0.54   &      \\
NGC 4314  &  0.56  &  0.31  & 160  &  21  &   Yes   & 1.04E$-2$   &  1.71   &  (d),(e)    \\
NGC 4321  &  0.87  &  0.32  & 170  &  51  &   Yes   & 6.64E$-3$   &  1.45   &      \\
NGC 5248  &  0.65  &  0.20  & 150  &  42  &   Yes   & 6.13E$-3$   &  1.23   &      \\
NGC 5728  &  1.10  &  0.23  & 180  &  40  &   Yes   & 1.09E$-2$   &  1.97   &      \\
NGC 5905  &  0.39  &  0.14  & 150  &  40  &     ?   & 3.88E$-3$   &  0.98   &      \\
NGC 5953  &  1.00  &  0.43  & 150  &  40  &     ?   & 9.50E$-3$   &  1.54   &      \\
NGC 6951  &  0.56  &  0.17  & 160  &  40  &   Yes   & 5.56E$-3$   &  1.25   &      \\
NGC 7552  &  0.34  &  0.15  & 150  &  40  &    No   & 3.38E$-3$   &  0.92   &  (f)    \\
NGC 7716  &  1.20  &  0.04  & 150  &  40  &    No   & 1.20E$-2$   &  1.72   &      \\
NGC IC14  &  0.68  &  0.36  & 204  &  40  &   Yes   & 6.57E$-3$   &  1.74   &
\enddata
\tablecomments{
Columns (2) and (3) give the semi-major axis and
ellipticity of nuclear rings adopted from \citet{com10}. Column (4) is the rotational velocity adopted from \citet{maz08} or from the
references given in Column (9). Column (5) is the total gas mass in the ring from \citet{she05} or from references in Column (9); we take $M_g = 4\times 10^8\Msun$ if no information is available. Column (6) cites the age distribution: ``Yes'' and ``No'' for the presence and absence of an azimuthal age gradient, respectively, and ``?'' for uncertain cases, adopted from \citet{all06}, \citet{maz08}, \citet{san10}, and \citet{bra12}.  Columns (7) and (8) give $\alpha$ and $\hatOmegas$ calculated by Equations \eqref{e:obs_alp} and \eqref{e:obs_Omg}. Column (9) is the references for $v_{\rm rot}$ or $M$: (a) \citet{oni15}; (b) \citet{hsi11}; (c) \citet{maz11}; (d) \citet{ gar91}; (e) \citet{ben96}; (f) \citet{bra12}.}
\end{deluxetable*}

The density distribution of finite slender rings obtained by our SCF
method for $\hatrB \lesssim -0.6$ is found to be well approximated by
Equation \eqref{e:rhosr}, which is also the solution for static,
isothermal cylinders of infinite extent. This indicates that the
rotation as well as geometrical curvature effect are insignificant in
determining an equilibrium for rings with the major axis $R_0$ much
longer than the minor axis $\eta_0$. The equilibrium angular frequency for isothermal slender rings with $\alpha\gtrsim 0.1$ is well described by Equation \eqref{e:Ostinc} applicable to truncated incompressible rings \citep{ost64b}.

To explore gravitational instability of nuclear rings, we calculate the growth rates of nonaxisymmetric modes with azimuthal mode number $m$ by assuming that the rings are slender with $\eta_0/R_0=0.1$, and that perturbations are independent of the polar angle $\lambda$ in the meridional plane. In the absence of rotation, the resulting dispersion relations are the same as those of axisymmetric modes for an infinite isothermal cylinder studied by \citet{nag87} if $m/R_0$ is taken equal to the wavenumber in the direction along the cylinder (see Fig.~\ref{f:rdisp_all}). Only large-scale modes can be gravitationally unstable, and the unstable range of $m$ as well as the maximum growth rate increase with decreasing $\alpha$.

Rotation tends to stabilize the gravitational instability, reducing both the growth rates and the unstable ranges of $m$.  The instability is completely suppressed when $\hatOmega$ exceeds the critical value that is relatively constant at $\sim 0.7$ for $\alpha\gtrsim0.01$ and increases rapidly with decreasing $\alpha$ (see Fig.~\ref{f:crit_cal}). The simple estimates of the critical angular frequencies from the Toomre condition as well as the local dispersion relation are smaller than the results of our full stability analysis for $\alpha \lesssim 5\times 10^{-3}$ due to underestimation of self-gravity at the ring centers. Shear turns out to be unimportant for the gravitational instability of rings as long as they are slender.

\subsection{Discussion}\label{s:dis}

\citet{maz08} analyzed photometric data of a sample of nuclear rings to estimate the ages of H$\alpha$-emitting star clusters and found that about a half of their sample contains an age gradient of star clusters along the azimuthal direction. Since nuclear rings with age gradient are thought to be gravitationally stable and form stars preferentially at the contact points, it is interesting to apply the results of our linear stability analysis to the observed rings to tell whether they are really stable.

Table \ref{t:obsring} lists the properties of 19 observed nuclear rings in galaxies with a noticeable bar, compiled from the literature where the information on the presence/absence of an age gradient is
available.\footnote{Most of the galaxies listed in Table
\ref{t:obsring} except for NGC 1097, NGC 2903, NGC 3351, and  NGC 7752 are adopted from \citet{maz08}: NGC 1097 is from \citet{san10}, NGC 2903 and NGC 3351 from \citet{maz11}, NGC 4321 from \citet{all06}, and NGC 7752 from \citet{bra12}.} Column (1) lists each galaxy name. Columns (2) and (3) give the semi-major axis and ellipticity of nuclear rings adopted from \citet{com10}.  Column (4) lists the rotational velocity $v_{\rm rot}$ adopted from \citet{maz08} or from the references given in Column (9). Column (5) lists the total gas mass $M_g$ in the ring from \citet{she05} or the references in Column (9) only for the galaxies with available data; we otherwise take $M_g=4\times10^8\Msun$ as a reference value. Column (6) indicates the presence or absence of an azimuthal age gradient of star clusters adopted from \citet{maz08}, \citet{all06}, \citet{san10}, and \citet{bra12}: a question mark is used when it is difficult to characterize the age distribution. Columns (7) and (8) give $\alpha$ and $\hatOmega$  calculated by
  \be\label{e:obs_alp}
    \alpha = 0.01 \left(\frac{c_s}{10\kms}\right)^2 \left(\frac{M_g}{4\times10^8\Msun}\right)^{-1}
 \left(\frac{R_0}{1\kpc}\right),
  \ee
and
  \be\label{e:obs_Omg}
 \hatOmega = 2.1 \left(\frac{v_{\rm rot}}{200\kms}\right) \left(\frac{M_g}{4\times10^8\Msun}\right)^{-0.5}
 \left(\frac{R_0}{1\kpc}\right)^{0.5},
  \ee
after taking $R_0=R_1(1-e^2)^{1/4}$ corresponding to the geometric means of the major and minor axes of eccentric rings, $\cs=10\kms$, and $\eta_0=R_0/10$. We replace $\rho_c$ with $\avgrho$ since the ring central density is difficult to constrain observationally.

\begin{figure}
\includegraphics[angle=0, width=8.5cm]{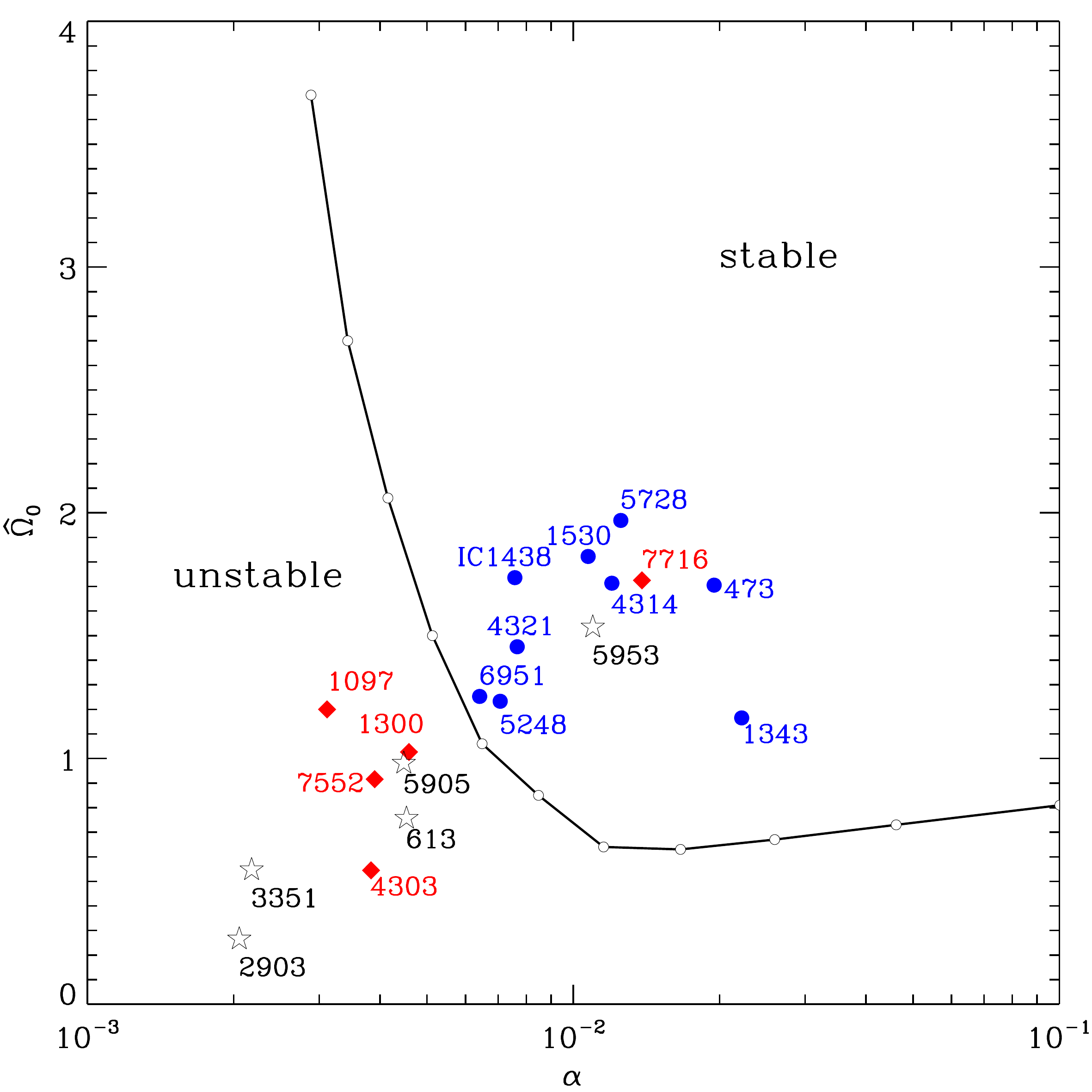}
\caption{Distributions of
$\alpha$ and $\hatOmega$ of the observed nuclear rings listed in Table \ref{t:obsring}. Blue circles and red diamonds represent rings with and without an azimuthal age gradient, respectively, while rings with uncertain age distributions are indicated by star symbols.
\label{f:crit_obs}}
\end{figure}

In Figure \ref{f:crit_obs}, we plot $\hatOmega$ against $\alpha$ for
the rings listed in Table \ref{t:obsring} using various symbols, with
numbers indicating galaxy names. Overall, rings with larger $\alpha$ tend to have larger $\hatOmega$. Blue circles represent rings with an azimuthal age gradient, while red diamonds are for those with no age gradient.  Rings for which the age distribution cannot be judged are marked by star symbols. It is apparent that all rings with an azimuthal age gradient are located at the stable regime, while all rings with no age gradient, except NGC 7716, correspond to unstable configurations. These results are consistent with two modes of star formation proposed by \citet{bok08}, such that rings sufficiently massive or rotating sufficiently slowly form stars in the popcorn style caused by gravitational instability, and thus do not show an apparent age gradient.  On the other hand, star formation in stable rings may occur preferentially at the contact points to exhibit an azimuthal age gradient of star clusters like pearls on a string.

The ring models we have considered so far ignored the effects of
magnetic fields that are pervasive in galaxies (e.g.,
\citealt{bec96,fle11}). In spiral galaxies, the presence of toroidal
magnetic fields is known to play a destabilizing role in forming giant clouds inside spiral arms where tension forces from bent field lines resist the stabilizing effect of the Coriolis force (e.g.,
\citealt{bal88,kim01,kim02,kim06}). In addition, magnetic fields
are likely to reduce the degree of central density concentration by
exerting pressure forces. It will be interesting to study how magnetic fields embedded in nuclear rings change the critical angular
frequencies for gravitational instability compared to those of
unmagnetized rings.

\acknowledgments

We acknowledge a helpful report from the referee, and are grateful to Dr.~Hsi-An Pan for the information on NGC 4321.  This work was supported by the National Research Foundation of Korea
(NRF) grant, No.~2008-0060544, funded by the Korea government (MSIP).
The computation of this work was supported by the Supercomputing
Center/Korea Institute of Science and Technology Information with
supercomputing resources including technical support (KSC-2015-C3-027).

\appendix

\section{Potential-density Pairs}\label{a:pairs}

\subsection{Spherical Coordinates}

For spheroid-like configurations, it is convenient to solve Equation \eqref{e:pos0} in spherical polar coordinates
$(r, \theta, \phi)$.
When an equilibrium is axially symmetric, we may expand the density using the Legendre function $P_l(\cos\theta)$ as
 \be\label{e:exp0}
    \rho (r, \theta ) =  \sum_{l=0}^\infty  \rho_l (r) P_l (\cos\theta),
 \ee
with the coefficients $\rho_{l}$ given by
 \be\label{e:exp1}
   \rho_l (r) = \frac{2l+1}{2} \int_0^\pi \rho (r, \theta ) P_l(\cos\theta)
   \sin \theta d\theta.
 \ee
Then, Equation \eqref{e:pos0} has the series solution of the form
  \be\label{e:pos1}
    \Phi(r,\theta) = -4\pi G  \sum_{l=0}^{l_{\rm max}}
    \frac{P_l (\cos\theta)}{2l +1}
    \left[\frac{1}{ r^{l+1}} \int_0^{ r} d r^\prime  r^{\prime l+1}  \rho_l (r^\prime) +
     r^l \int_{r}^\infty \frac{d r^\prime }{r^{\prime l-1}} \rho_l (r^\prime)       \right],
   \ee
(see, e.g., Eq. (2.95) of \citealt{bin08,sir09}). The summation in Equation \eqref{e:pos1} is truncated typically at $l_{\rm max}=50$ in our computation.

\subsection{Toroidal Coordinates}

The multipole expansion using Legendre functions described above becomes inefficient for highly flattened systems such as slender rings/tori because it requires a high cutoff number $l_\mathrm{max}$ for accurate potential evaluation.  To resolve such equilibria, the SCF method in the spherical coordinates also needs a very large computation grid.  If we instead employ toroidal coordinates, flattened equilibria are well resolved with a relatively small grid and a very small number of multipole terms.

The toroidal coordinates ($\tau$, $\sigma$, $\phi$) are defined by
  \be\label{e:toroidal}
    \left(
  \begin{array}{c}
     x \\ y \\ z
  \end{array} \right)
   = \left(
  \begin{array}{c}
    a\sinh\sigma\cos\phi   \\
    a\sinh\sigma\sin\phi     \\
    a\sin\tau
  \end{array} \right) \Bigg/ f(\sigma,\tau),
 \ee
where $a$ is a constant (focal length) and
  \be
    f(\sigma,\tau)\equiv \cosh\sigma-\cos\tau.
  \ee
Note that $\sigma\in[0, \infty)$ is a (dimensionless) radial distance, $\tau\in[0, 2\pi)$ is a poloidal angle, and $\phi\in[0, 2\pi)$ is equal to the usual cylindrical azimuthal angle. The constant-$\sigma$ surface is a torus with a circular cross section: it becomes a focal ring with radius $a$ when $\sigma = \infty$.

In this coordinate system, the Green's function for a Laplacian operater can be expanded based on the half-integer degree Legendre functions $P_{l-1/2}^m$ and $Q_{l-1/2}^m$ as
\begin{equation}\label{eq:toroidalGreen}
\begin{split}
    \frac{1}{|\mathbf{x}-\mathbf{x}^\prime|} = \frac{(ff^\prime)^{1/2}}{a\pi}
     \sum_{l,m=0}^\infty \epsilon_l\epsilon_m (-i)^m \frac{\Gamma(l-m+\tfrac{1}{2})}{\Gamma(l+m+\tfrac{1}{2})}
     \cos[l(\tau-\tau^\prime)] \cos[m(\phi-\phi^\prime)]  \\ \times
    \begin{cases}
        P_{l-{1}/{2}}^m(\cosh\sigma) Q_{l-{1}/{2}}^m(\cosh\sigma^\prime),\;\;\;\textrm{for}\; &\sigma^\prime > \sigma, \\
        P_{l-{1}/{2}}^m(\cosh\sigma^\prime) Q_{l-{1}/{2}}^m(\cosh\sigma),\;\;\;\textrm{for}\; &\sigma^\prime < \sigma,
    \end{cases}
\end{split}
\end{equation}
where $f^\prime\equiv f(\sigma^\prime, \tau^\prime)$, and
$\epsilon_k$ is equal to unity for $k=1$ and two otherwise
(e.g., \citealt{mor53,coh99,coh00}).\footnote{As noted by \citet{coh00}, there are some typographical mistakes in Equation (10.3.81) of \citet{mor53}: the focal length $a$ is missing and a sign in the argument of a Gamma function is incorrect.}

The gravitational potential is then given by
\begin{equation}\label{e:tG1}
\begin{split}
    \Phi(\sigma,\tau) &= - G \int_V \frac{\rho(\mathbf{x}^\prime)}{|\mathbf{x}-\mathbf{x}^\prime|} d^3x^\prime\\
    &= - G\int_0^{2\pi}d\phi^\prime
         \int_0^{2\pi}d\tau^\prime
         \int_0^\infty d\sigma^\prime
    \frac{\rho(\mathbf{x}^\prime)}{|\mathbf{x}-\mathbf{x}^\prime|} \frac{a^3\sinh\sigma^\prime}{f(\sigma^\prime,\tau^\prime)^3}.
\end{split}
\end{equation}
For an axisymmetric mass distribution $\rho(\sigma,\tau)$, Equation \eqref{e:tG1} reduces to
\begin{equation}\label{e:gpot_toroidal}
\begin{split}
    \frac{\Phi(\sigma,\tau)}{4a^2Gf(\sigma,\tau)^{1/2}}
    = &- \sum_{l=0}^{l_\mathrm{max}}\epsilon_l P_{l-1/2}(\cosh\sigma)\cos(l\tau) \int_\infty^\sigma d\sigma^\prime \sinh\sigma^\prime Q_{l-1/2}(\cosh\sigma^\prime) \rho_l(\sigma^\prime) \\
    &- \sum_{l=0}^{l_\mathrm{max}} \epsilon_l Q_{l-1/2}(\cosh\sigma)\cos(l\tau) \int_0^\sigma d\sigma^\prime \sinh\sigma' P_{l-1/2}(\cosh\sigma^\prime) \rho_l(\sigma^\prime),
\end{split}
\end{equation}
where
 \be\label{e:rho_to}
 \rho_l (\sigma^\prime) = \int_0^\pi \frac{\rho(\sigma^\prime,\tau^\prime)\cos(l\tau^\prime)}{f(\sigma^\prime,\tau^\prime)^{5/2}} d\tau^\prime,
 \ee
and we assume a reflection symmetry of $\rho(\sigma,\tau)$ with respect to the $\tau=0$ plane.

\subsection{Comparison}\label{a:comp}

\begin{figure*}
\includegraphics[angle=0, width=18cm]{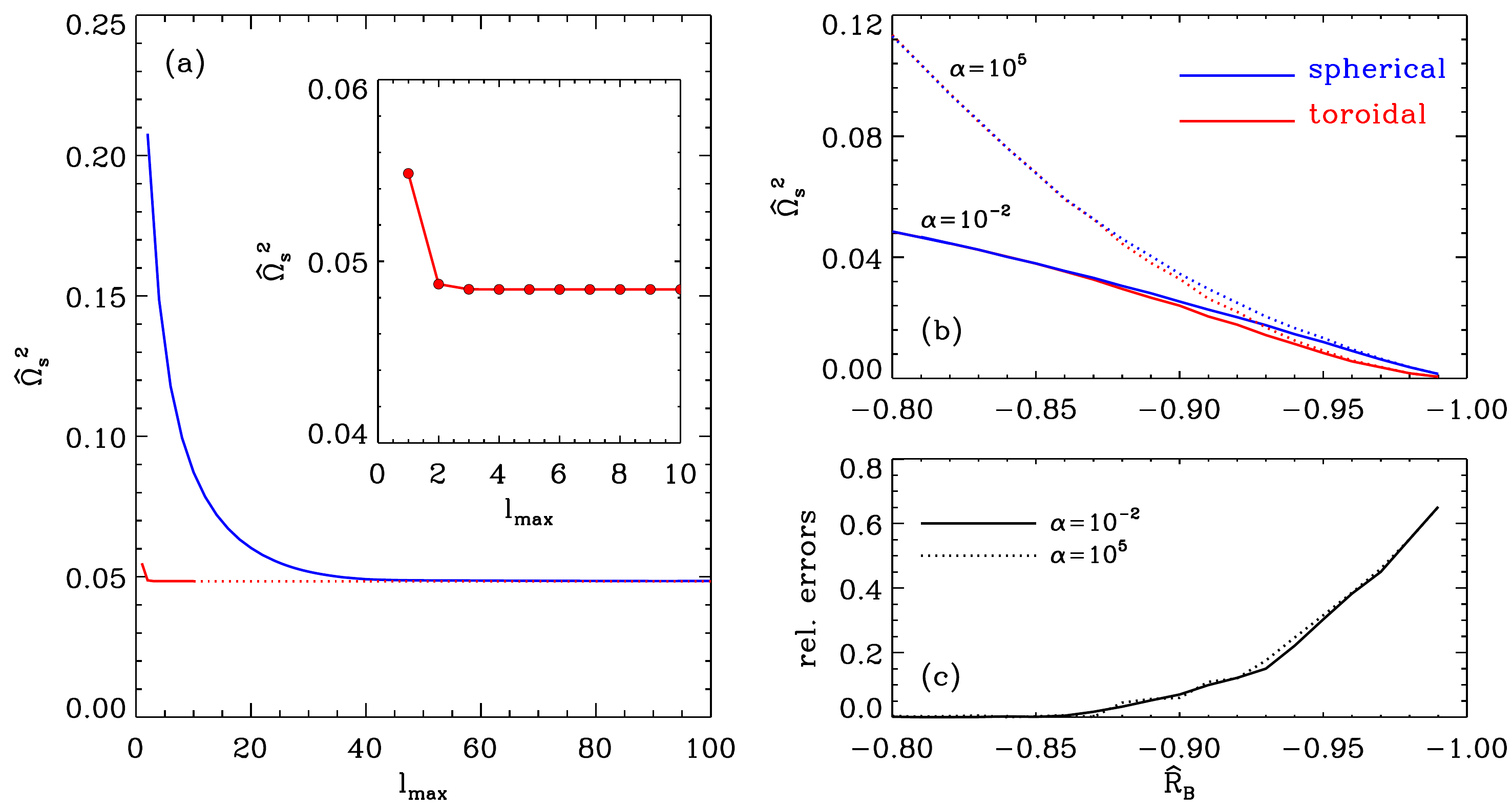}
\caption{
(a): Convergence of $\hatOmegas^2$ against $l_\mathrm{max}$ for a ring-like equilibrium with $\hatrB=-0.8$ and $\alpha=0.01$. The blue and red solid lines draw $\hatOmegas^2$ based on the spherical and toroidal multipole expansions, respectively. The red dotted line at $l_{\rm max}\geq10$ is the extrapolation of the result with $l_{\rm max}=10$. The inset zooms in the toroidal results with $l_{\rm max}\leq10$. (b): Comparison between $\hatOmegas^2$  from the spherical expansions with $l_{\rm max} = 50$ and the toroidal expansions with $l_{\rm max} = 10$  for ring-like equilibria with $-0.80\geq\hatrB\geq -0.99$ with
$\alpha=10^{-2}$ (solid lines) or $\alpha=10^5$ (dotted lines). (c): Relative errors, $\hatOmegas^2(\text{sph.})/\hatOmegas^2(\text{tor.})-1$, between the two methods.
\label{f:comp}}
\vspace{0.2cm}
\end{figure*}

To compare the rate of convergence against $l_{\rm max}$ as well as accuracy between the results from the two different multipole expansions, we calculate ring-like equilibria with $\hatrB\leq -0.8$ on both spherical and toroidal meshes.  Figure \ref{f:comp}(a) plots $\hatOmegas^2$ as a function of $l_{\rm max}$ for an equilibrium with $\hatrB=-0.8$ and $\alpha=0.01$. The blue and red lines correspond to the results with the spherical and toroidal multipole expansions, respectively. The inset zooms in the results with the toroidal mesh, showing that the SCF method converges to $\hatOmega^2=0.048$ very rapidly when the potential is expanded in the toroidal coordinates.  On the other hand, the SCF method with the spherical multipole expansions converges quite slowly, requiring $l_{\rm max} \gtrsim 42$ to obtain the solution within 1\% of the value from the toroidal multipole expansion shown as the dotted line.

Figure \ref{f:comp}(b) compares $\hatOmegas^2$ obtained by
the spherical mutipole expansion with $l_{\rm max}=50$ and
the toroidal mutipole expansion with $l_{\rm max}=10$  for ring-like equilibria with $-0.80\geq \hatrB\geq -0.99$ and $\alpha=10^{-2}$ (solid lines) and $\alpha=10^5$ (dotted lines). The relative errors given in Figure \ref{f:comp}(c) show that the two methods agree within 1\% for $\hatrB \gtrsim -0.86$, and the spherical expansions overestimate $\hatOmegas^2$ for smaller $\hatrB$. This is mainly because such flattened equilibria are not well resolved in the spherical mesh. For this reason, we employ the toroidal potential expansion in constructing equilibria with $\hatrB \leq -0.8$ presented in Section \ref{s:seq}.

\section{Dynamical Equations for Slender Rings}\label{a:eqn}

In this Appendix, we provide expressions for gas dynamical equations
for slender rings in the new curvilinear coordinates $(\eta, \lambda,
\phi)$. It is straightforward to show that differentiations of $x$,
$y$, and $z$ in Equation \eqref{e:coord} yield
  \be
   dx^2 + dy^2 + dz^2 = d\eta^2 + \eta^2 d\lambda^2 + (R_0 + \eta \cos\lambda)^2 d\phi^2,
  \ee
so that the scale factors in the $\eta$-, $\lambda$-, and
$\phi$-directions are given by
  \be
  h_\eta = 1,\;\;\; h_\lambda=\eta,\;\;\textrm{and}\;\; h_\phi= R_0 + \eta \cos\lambda,
  \ee
respectively.

The unit vectors in the curvilinear coordinates can be resolved into
their Cartesian components as
  \be\label{e:uv}
  \begin{split}
  {\bf e}_\eta    &=  {\bf i}\cos\lambda \cos\phi  + {\bf j}\cos\lambda \sin\phi  + {\bf k}\sin\lambda , \\
  {\bf e}_\lambda &= -{\bf i}\sin\lambda \cos\phi  - {\bf j}\sin\lambda \sin\phi  + {\bf k}\cos\lambda , \\
  {\bf e}_\phi    &= -{\bf i}\sin\phi  + {\bf j}\cos\phi,
 \end{split}
 \ee
where ${\bf i}$, ${\bf j}$, and ${\bf k}$ refers to the unit vectors in the $x$-, $y$-, and $z$-directions. It is clear that the unit vectors are orthogonal to each other. The angular derivatives of the unit vectors are
 \be\label{e:duv1}
   \frac{\partial {\bf e}_\eta}{\partial \lambda} = {\bf e}_\lambda,\;\;\;
   \frac{\partial {\bf e}_\eta}{\partial \phi} = {\bf e}_\phi\cos\lambda ,\;\;\;
 \ee
  \be\label{e:duv2}
   \frac{\partial {\bf e}_\lambda}{\partial \lambda} = - {\bf e}_\eta,\;\;\;
   \frac{\partial {\bf e}_\lambda}{\partial \phi} = -{\bf e}_\phi\sin\lambda ,\;\;\;
 \ee
 \be\label{e:duv3}
   \frac{\partial {\bf e}_\phi}{\partial \lambda} = 0,\;\;\;
   \frac{\partial {\bf e}_\phi}{\partial \phi} = - {\bf e}_\eta\cos\lambda  + {\bf e}_\lambda\sin\lambda.
 \ee

The gradient of any scalar field $\psi$ in the new coordinates is given by
  \be\label{e:gra}
    \nabla\psi = {\bf e}_\eta \frac{\partial \psi}{\partial \eta} +
                 {\bf e}_\lambda \frac{1}{\eta} \frac{\partial \psi}{\partial \lambda}+
                 {\bf e}_\phi \frac{1}{1 + (\eta/R_0) \cos\lambda} \frac{1}{R_0} \frac{\partial \psi}{\partial \phi},
  \ee
while the divergence of an arbitrary vector field
${\bf V} = V_\eta {\bf e}_\eta + V_\lambda {\bf e}_\lambda + V_\phi {\bf e}_\phi$
is
 \be\label{e:div}
 \begin{split}
 \nabla\cdot {\bf V} &=
   \frac{\partial V_\eta}{\partial \eta}
   +  \frac{1+2(\eta/R_0)\cos\lambda}{1+(\eta/R_0)\cos\lambda} \frac{V_\eta}{\eta}
   +  \frac{1}{\eta} \frac{\partial V_\lambda}{\partial \lambda}   \\
   &-    \frac{\sin\lambda}{1+(\eta/R_0)\cos\lambda} \frac{V_\lambda}{R_0}
   + \frac{1}{1+(\eta/R_0)\cos\lambda}  \frac{1}{R_0} \frac{\partial V_\phi}{\partial \phi}.
  \end{split}
 \ee
By taking ${\bf V} = \nabla \psi$, one can obtain an expression for the Laplacian of $\psi$ as
 \be\label{e:lap}
 \begin{split}
   \nabla^2 \psi & =
    \frac{\partial^2\psi}{\partial \eta^2} +
    \frac{1+2(\eta/R_0)\cos\lambda}{1+(\eta/R_0)\cos\lambda} \frac{1}{\eta}\frac{\partial \psi}{\partial \eta} +
    \frac{1}{\eta^2} \frac{\partial^2 \psi}{\partial \lambda^2} \\
    &-    \frac{\sin\lambda}{1+(\eta/R_0)\cos\lambda} \frac{1}{R_0 \eta} \frac{\partial \psi}{\partial \lambda} +
    \frac{1}{[1+(\eta/R_0)\cos\lambda]^2}  \frac{1}{R_0^2} \frac{\partial^2\psi}{\partial \phi^2},
  \end{split}
 \ee
(e.g., \citealt{ost64b}).

Under the slender-ring approximation of $\eta/R_0 \ll1$,
Equations \eqref{e:pos0},
\eqref{e:con}, and \eqref{e:mom} are reduced to
 \be\label{e:con2}
   \frac{\partial \rho}{\partial t} + \frac{1}{\eta}\frac{\partial (\eta \rho v_\eta)}{\partial \eta}
    + \frac{1}{\eta}\frac{\partial (\rho v_\lambda)}{\partial \lambda} - \sin\lambda \frac{\rho v_\lambda}{R_0}
    + \frac{1}{R_0} \frac{\partial (\rho v_\phi) }{\partial \phi} =0,
 \ee
 \be\label{e:meta2}
   \frac{\partial v_\eta}{\partial t} +
    {\bf v}\cdot \nabla v_\eta
    - \cos\lambda \frac{v_\phi^2}{R_0} - \frac{v_\lambda^2}{\eta}
     = -\frac{\cs^2}{\rho}\frac{\partial \rho}{\partial \eta}
       -  \Omega_e^2  R_0 \cos\lambda -  \frac{\partial \Phi_s}{\partial \eta}
 \ee
  \be\label{e:mlam2}
  \frac{\partial v_\lambda}{\partial t} +
   {\bf v}\cdot \nabla v_\lambda
   +  \frac{v_\eta v_\lambda}{\eta}
   + \sin\lambda \frac{v_\phi^2}{R_0}
    = -\frac{\cs^2}{\rho\eta}\frac{\partial \rho}{\partial \lambda}
     +  \Omega_e^2 R_0 \sin\lambda
     - \frac{1}{\eta} \frac{\partial \Phi_s}{\partial \lambda}
 \ee
 \be\label{e:mphi2}
  \frac{\partial v_\phi}{\partial t} +
 {\bf v}\cdot \nabla v_\phi
   - \sin\lambda \frac{v_\lambda v_\phi}{\eta}
   + \cos\lambda \frac{v_\eta v_\phi}{R_0}
    = -\frac{\cs^2}{\rho R_0}\frac{\partial \rho}{\partial \phi}  -
    \frac{1}{R_0} \frac{\partial \Phi_s}{\partial \phi}
 \ee
 \be\label{e:pos2}
  \frac{\partial^2\Phi_s}{\partial \eta^2}
   + \frac{1}{\eta} \frac{\partial\Phi_s}{\partial \eta}
   + \frac{1}{\eta^2}\frac{\partial\Phi_s^2}{\partial \lambda^2}
   - \frac{\sin\lambda}{R_0\eta} \frac{\partial \Phi_s}{\partial \lambda}
     + \frac{1}{R_0^2} \frac{\partial^2\Phi_s}{\partial \phi^2} = 4\pi G\rho.
 \ee
where
  \be\label{e:vdel}
    {\bf v}\cdot \nabla =  v_\eta \frac{\partial}{\partial \eta} +
      \frac{v_\lambda}{\eta}\frac{\partial }{\partial \lambda} +
     \frac{v_\phi}{R_0}\frac{\partial}{\partial \phi}.
  \ee
These can be further simplified to Equations
\eqref{e:con3}--\eqref{e:pos3} under the assumptions that the fluid
variables are independent of $\lambda$, and that
$v_\lambda=\sin\lambda=0$.

\section{Local Dispersion Relation}\label{a:wkb}

Here we derive a local dispersion relation for waves with $m/R_0 \gg |d \ln (\eta\rho_0)/d\eta|$. We assume that $\chi_1$ does not vary much with $\eta$ (i.e., $|d\ln \chi_1/d\eta| \ll m/R_0$) as well, which is usually the case for gravitationally unstable modes. Equation \eqref{e:lcon2} then reduces to
 \be\label{e:w1}
        \left(\tomega^2-4\Omega_0^2 - c_s^2\frac{m^2}{R_0^2}\right) \frac{\rho_1}{\rho_0} = \frac{m^2}{R_0^2} \Phi_{s1}.
 \ee

Using the Green's function method, it is straightforward to find a
formal solution of Equation \eqref{e:lpos1} as
 \be\label{e:g0}
   -\frac{\Phi_{s1}}{4\pi G} = K_0(x)\int_0^x I_0(x^\prime) \rho_1(x^\prime) x^\prime dx^\prime + I_0 (x) \int_x^\infty K_0 (x^\prime) \rho_1 (x^\prime) x^\prime dx^\prime,
 \ee
where $x\equiv (m/R_0)\eta$ and $K_0$ and $I_0$ are the modified Bessel functions. Since the perturbations are assumed insensitive to $\eta$, we may take
 \be\label{e:pden0}
    \rho_1(x) =
     \begin{cases}
     \text{constant}  ,\;\; &\textrm{for} \;x \leq x_0=(m/R_0)\eta_0, \\
      0,   \;\; &\textrm{otherwise}.
  \end{cases}
  \ee
Then, Equation \eqref{e:g0} yields
  \be\label{e:g1}
  -\frac{\Phi_{s1}}{4\pi G\rho_1} = x[ K_0(x)I_1(x) + I_0(x)K_1(x)] - x_0 K_1(x_0) I_0(x),
  \ee
where [G5.56.2] is used.\footnote{[G$\cdots$] refers to the formula number in the integral table of \citet{gra94}.} Using [G8.477.2], Equation \eqref{e:g1} is further simplified to
  \be\label{e:g2}
    \Phi_{s1} = -4\pi G\rho_1 [ 1 -  x_0 K_1(x_0) I_0(x) ].
  \ee

Taking the perturbed gravitational potential at the center ($x=0$) of the ring, and plugging Equation \eqref{e:g2} into Equation \eqref{e:w1}, we obtain the local dispersion relation
  \be\label{e:ldisp}
    \tomega^2 = 4\Omega_0^2
  - 4\pi G\rho_0 \left[ 1 -  \frac{m\eta_0}{R_0} K_1\left( \frac{m\eta_0}{R_0} \right) \right]+ c_s^2\frac{m^2}{R_0^2}.
  \ee
Note that the second terms in the right-hand side of Equation \eqref{e:ldisp} is equal to
the gravity term in Equation (14) of \citet{elm94} if $(m/R_0)\eta_0$ is replaced by $k\Delta r$.

\end{document}